\documentclass[acmsmall]{acmart}
\AtBeginDocument{%
  }

\setcopyright{cc}
\setcctype{by}
\acmDOI{10.1145/3715740}
\acmYear{2025}
\acmJournal{PACMSE}
\acmVolume{2}
\acmNumber{FSE}
\acmArticle{FSE025}
\acmMonth{7}
\received{2024-09-13}
\received[accepted]{2025-01-14}

\usepackage{comment}
\usepackage[utf8]{inputenc}
\usepackage{hyperref}
\usepackage{algorithmic} 
\usepackage{graphicx}
\usepackage{textcomp}
\usepackage{tikz}
\usepackage{xcolor}
\usepackage{xspace}
\usepackage[ruled,linesnumbered]{algorithm2e}
\usepackage{colortbl}
\usepackage{mdframed}
\usepackage{subcaption}
\usepackage{svg}
\usepackage{wrapfig}

\usepackage{multirow}
\usepackage{booktabs}
\usepackage[flushleft]{threeparttable}
\usepackage[most]{tcolorbox}
\usepackage{rotating}
\usepackage[normalem]{ulem}
\useunder{\uline}{\ul}{}
\newcommand*\rot{\rotatebox{90}}

\definecolor{myred}{RGB}{214,39,40}
\definecolor{mygreen}{RGB}{44,160,44}
\definecolor{myblue}{RGB}{32,119,180}
\definecolor{myorange}{RGB}{255,127,14}
\definecolor{mygrey}{HTML}{EFEFEF}

\newboolean{showcomments}
\setboolean{showcomments}{false} 
\ifthenelse{\boolean{showcomments}}
  {\newcommand{\nb}[2]{
  \fbox{\bfseries\sffamily\scriptsize#1}
     {\sf\small$\blacktriangleright$\textit{\textcolor{red}{#2}}$\blacktriangleleft$}
   }
  }
  {\newcommand{\nb}[2]{}
   
  }

\newcommand\matteo[1]{\nb{Matteo}{#1}}

\newcommand{\head}[1]{\noindent\textbf{#1.}}

\newcommand{\art}{\textsc{ART}\xspace}
\newcommand{\rand}{\textsc{Rand}\xspace}
\newcommand{\dist}{\textsc{Dist}\xspace}
\newcommand{\gram}{\new{\textsc{q-gram}}\xspace}
\newcommand{\grams}{\new{\textsc{q-grams}}\xspace}
\newcommand{\gramss}{\new{\textsc{q-grams\textsubscript{s}}}\xspace}
\newcommand{\gramssi}{\new{\textsc{q-grams\textsubscript{s+i}}}\xspace}

\newcommand{\new}[1]{\textcolor{black}{#1}}

\newcommand{\colorline}[1]{\raisebox{0.5ex}{\textcolor{#1}{\rule{0.5cm}{2pt}}}}

\begin{document}

\title{Adaptive Random Testing with Q-grams: The Illusion Comes True}

\author{Matteo Biagiola}
\orcid{0000-0002-7825-3409}
\affiliation{%
	\institution{Università della Svizzera italiana}
	\city{Lugano}
	\country{Switzerland}
}
\email{matteo.biagiola@usi.ch}

\author{Robert Feldt}
\orcid{0000-0002-5179-4205}
\affiliation{%
	\institution{Chalmers University of Technology}
	\city{Gothenburg}
	\country{Sweden}
}
\email{robert.feldt@chalmers.se}

\author{Paolo Tonella}
\orcid{0000-0003-3088-0339}
\affiliation{%
	\institution{Università della Svizzera italiana}
	\city{Lugano}
	\country{Switzerland}
}
\email{paolo.tonella@usi.ch}

\begin{abstract}
  Adaptive Random Testing (\art) has faced criticism, particularly for its computational inefficiency, as highlighted by Arcuri and Briand.  
Their analysis clarified how  \art requires a quadratic number of distance computations as the number of test executions increases, which limits its scalability in scenarios requiring extensive testing to uncover faults.
Simulation results support this, showing that the computational overhead of these distance calculations often outweighs \art's benefits. While various \art variants have attempted to reduce these costs, they frequently do so at the expense of fault detection, lack complexity guarantees, or are restricted to specific input types, such as numerical or discrete data.

In this paper, we introduce a novel framework for adaptive random testing that replaces pairwise distance computations with a compact aggregation of past executions, such as counting the \grams observed in previous runs. 
Test case selection then leverages this aggregated data to measure diversity (e.g., entropy of \grams), allowing us to reduce the computational complexity from quadratic to linear.

Experiments with a benchmark of six web applications, show that \art with \grams covers, on average, $4 \times$ more unique targets than random testing, and $3.5 \times$ more than \art using traditional distance-based methods.

\end{abstract}



\received{12 September 2024}

\maketitle

\section{Introduction} \label{sec:intro}

Adaptive Random Testing (ART)~\cite{ChenLM04} operates on the assumption that failure regions in the input space are contiguous, separated by large areas where the software performs without error. Based on this, a random test generator can be optimized by ensuring broader exploration of the input space, by increasing diversity. \art achieves this by first generating a set of candidate inputs and then selecting those at the farthest distance from previously executed inputs.

While \art enhances the exploration of the input space, it introduces a significant computational cost: calculating the distances between all new candidates and every previously executed test.  This observation led Arcuri and Briand~\cite{ArcuriB11} to deem \art's effectiveness an ``illusion of effectiveness''. In their work, they performed a rigorous theoretical analysis, further supported by simulations, demonstrating that the benefits of \art's reduction in the number of test executions--achieved through a higher probability of failure detection--are outweighed by the computational expense of distance calculations in nearly all practical scenarios. 
Specifically, when a fault leads to a failure with only a low probability, making it difficult to detect, \art's additional computational costs only pay off when the test execution time is unrealistically longer than the cost of distance computation.

In this paper, we introduce a new framework for adaptive random testing that removes the need for pairwise distance calculations, reducing ART’s computational cost from quadratic to linear. The key innovation lies in how we manage previously executed tests: we compute an intermediate structure, called the aggregation, which is easily updated incrementally.
Rather than comparing each new input to all prior tests individually, the distance is calculated once against the aggregation. We instantiate this framework using \grams, consecutive sequences of $Q$ tokens in the input, where the \gram counts of the executed set serves as the aggregation. This structure allows for easy updates as new inputs are selected or considered. To measure diversity, we can calculate metrics like entropy from \gram counts.

We replicated the theoretical analysis conducted by Arcuri and Briand~\cite{ArcuriB11} and demonstrated that \gram aggregation indeed reduces the complexity of diversity computations from quadratic to linear. As a consequence, for ART to outperform random testing, the test execution time need only be one order of magnitude (or more, naturally) greater than the time for entropy computation. We validated these theoretical insights through a simulation experiment modeled after Arcuri and Briand’s work. The results show that, in low-probability fault scenarios, \art with \gram aggregation is significantly more effective than random testing, exposing faults with a 2.6$\times$ to 3.4$\times$ higher probability and requiring approximately 10$\times$ fewer test executions. In contrast, ART with traditional distance computation proved impractical.

We further applied our \grams-based \art test selector to six web applications, for which we found that test execution times were three orders of magnitude higher than the cost of entropy computation (i.e. measured in seconds versus milliseconds). The experimental results reveal that \art with \gram aggregation not only achieves greater functional coverage of the web applications but also improves efficiency and identifies more unique coverage targets compared to both random testing and \art with distance computation. This advantage is especially pronounced in web applications with hard-to-cover functionalities. Across all web applications tested, \art with \gram aggregation covers, on average, 4$\times$ more unique targets than random generation and 3.5$\times$ more than ART with distance computation.

\section{Related Work} \label{sec:rel-work}

\subsection{Adaptive Random Testing}

Since its initial formulation~\cite{ChenLM04}, numerous variants of Adaptive Random Testing~\cite{ChenLM04} (\art) have been developed. These approaches primarily focus on either partitioning the input space to enable random sampling within each partition~\cite{chen2004adaptive,shahbazi2013centroidal}, or minimizing the number of distance computations required to find the nearest element. The latter is accomplished through techniques such as ``forgetting'' prior selections~\cite{chan2006forgetting}, or by structuring the selections more efficiently within appropriate data structures~\cite{MaoZTC19}. \new{Due to page limitations, in the following we discuss examples of different ``families'' of approaches. Specifically, approaches that use specialized data structures to reduce the number of distance computations (to \textit{O(N $\cdot$ $\log$(N))})~\cite{MaoZTC19, shahbazi2013centroidal}, and approaches that present a linear-time algorithm~\cite{barus2016cost,shahbazi2013centroidal} that can also handle non-numeric inputs~\cite{barus2016cost}.}

A recent and notable example of the latter is KDFC-ART~\cite{MaoZTC19} that attempts to mitigate the quadratic cost of distance computation in \art by taking advantage of $k$-dimensional trees (KD-trees)~\cite{Bentley75}. Instead of computing the distance between each candidate test and all previously executed tests, a KD-tree is traversed to find the subset of previously executed tests in the nearest neighbour. However, there is no guarantee that the test at minimum distance belongs to the KD-tree node representing the nearest neighbour. Consequently, the algorithm may need to backtrack, and in the worst-case scenario (a full backtrack to the tree root), the computational cost remains quadratic. Additionally, this method, as the majority of \art variants, is limited to fixed-size numeric vectors, excluding programs that handle strings or variable-sized vectors/data structures. \new{Similarly Ackah-Arthur et al.~\cite{ackah2019onedomain} proposed a bisection method that require a similar number of test executions to other ART methods while lowering the test input selection costs via a binary tree structure. However, it is unclear how this approach can be used for non-numeric inputs, since it requires a way to split/sub-divide the input space dimensions.}

Barus et al~\cite{barus2016cost} introduced the ArtSum algorithm, a linear-order \art method for software with non-numeric inputs. ArtSum uses a distance measure based on manually pre-selected categories of inputs, treating any two different choices for a category as distinct without considering their degree of difference.  This approach allows the distance of a new candidate to all previously selected inputs to be computed in linear time, significantly enhancing computational efficiency. However, this method requires the identification of suitable categories and choices, which may demand extensive domain knowledge and effort from testers. Additionally, it is unclear how to handle inputs containing both numeric and discrete values. \new{The approach by Shahbazi et al.~\cite{shahbazi2013centroidal} is a diversity-based linear-order \art method, but only considers numerical inputs, spreading them uniformly across the input space.}

Compared to existing methods for fixed-size candidate-set \art, our approach is applicable to combinations of inputs or test steps of any type (as they can be serialized to strings), operates in linear time, and ensures that updates occur only when a formally defined diversity measure, such as entropy, is improved.

\subsection{Diversity-based Test Generation and Selection} 

Elgendy et al. provide a comprehensive survey of diversity-based techniques in software testing, identifying over 70 distinct metrics, with the most common artifacts being test scripts or test inputs~\cite{Elgendy2023}. While many distance metrics\footnote{Referred to as similarity metrics in Elgendy et al.'s survey; however, similarity metrics can typically be converted into distance metrics, so we use the latter term here.} are specific to data types (e.g., extended subtree metrics~\cite{shahbazi2014extended}) or tailored to particular applications or approaches, several are general enough to be applicable across various data types or scenarios.

A recent empirical study comparing various general diversity metrics for selecting small subsets from large, automatically generated test suites~\cite{elgendy2024evaluating}, found that the normalized compression distance, first proposed for software testing by Feldt et al.~\cite{feldt2008searching}, was the most effective for fault detection. However, compression is often computationally expensive, particularly when scaled to a set of objects, such as the Test Set Diameter, which has a quadratic cost relative to the test set size~\cite{feldt2016tsdm}.

Entropy offers a fast and flexible measure for quantifying information uncertainty based on the distribution of counts or frequencies of objects or symbols~\cite{renyi1961measures}, with its intrinsic link to compression and diversity measures often highlighted~\cite{sherwin2010entropy}. Shimari et al.~\cite{shi2015measuring} used entropy as a diversity metric by evaluating the distribution of test case distances, showing that this ``distance entropy'' outperformed simpler diversity metrics without added computational complexity. Another method for applying entropy in test case diversity involves \gram language models, as demonstrated by Leveau et al.~\cite{leveau2022fostering}, who used this technique to increase the diversity of exploratory testers and improve their fault-detection abilities in web application testing. In the same domain, Biagiola et al.~\cite{Biagiola0RT19} employed sequence edit distance to ensure diversity among web test cases.

\section{Background on Adaptive Random Testing} \label{sec:background}

\subsection{Algorithm}

\begin{algorithm}[t]
\caption{\art with pairwise distance computation (\dist) (pseudocode adapted from Arcuri and Briand~\cite{ArcuriB11})} \label{algo:art-dist}

$Z$ = \{\} \;
\textit{add a random test case to $Z$ and execute it} \;
\While{\textit{stopping criterion not satisfied}}{
    \textit{sample set $W$ of random test cases} \;
    \ForEach{$w \in W$}{
        $w$.\textit{minD} = min(dist($w$, $z \in Z$)) \;
    }
    $w^*$ = $\arg\max_{w \in W}$ \{ \textit{$w$.minD} \} \;
    \textit{execute $w^*$ and add it to $Z$} \;
}

\end{algorithm}

\autoref{algo:art-dist} shows the pseudocode describing the general test generation procedure behind Adaptive Random Testing (\art)~\cite{ChenLM04}. The key difference with respect to random testing (\rand) is that \art evaluates a set of randomly generated candidates ($W$ at Line~4) for their diversity w.r.t. the previously executed tests. Diversity is assessed as the distance between the candidates (set $W$) and the archive $Z$ of previously executed tests (Lines~5---7). The test $w^*$ with maximum distance\footnote{Actually, the maxi-min, i.e. the largest minimum distance, is typically used even though other choices are also possible.} from the archive among the candidates is executed and added to the archive (Lines~8--9). The specific distance metric used to quantify diversity is problem dependent and can be for instance Euclidean distance when the test input is a tuple of numbers or string edit distance if the input is a string.

At each iteration of the main loop, \autoref{algo:art-dist} performs $|W| \times |Z|$ distance computations. As shown formally in previous work~\cite{ArcuriB11}, if the algorithm stops after executing $\rho$ test cases, the number of distance computations it requires is quadratic with $\rho$: $\sum_{i=1}^{\rho-1} |W|i = |W|\rho(\rho-1) = \Theta(\rho^2)$. 

\subsection{Illusion of Effectiveness}

In their paper~\cite{ArcuriB11}, Arcuri and Briand point out that the quadratic cost of distance computations in \art will diminish the time available for executing test cases, potentially negating the benefits ART is supposed to bring over random testing (\rand). They conducted repeated simulation experiments on the triangle classification program, focusing on a mutant with an estimated failure probability of $\theta = 1.51 \cdot 10^{-5}$. Their results showed no statistically significant difference between \art and \rand in the number of test cases needed to be sampled. However, they found a significant difference in execution time: \art took an average of 47.7 minutes to expose the fault (kill the mutant), while \rand required only 10.3 milliseconds. This stark contrast is largely attributable to the substantial overhead introduced by distance computations in \art.

Arcuri and Briand, based on the results of their simulation experiment, then analyzed why empirical evidence appears to favor \art over \rand, describing this as an ``illusion of effectiveness''. They arrived at several key conclusions: (1) existing studies relied on unrealistic assumptions (i.e., very high failure rates ($\theta$)); (2) these studies examined only a small set of programs, with faults introduced in an unsystematic manner (i.e., failing to ensure low $\theta$); (3) most studies focused solely on the number of test executions needed to expose a fault, neglecting the time \art requires for distance computation, which \rand can use to detect faults sooner; and (4) automated oracles (except for simple crash oracles) were rarely employed.

The authors~\cite{ArcuriB11} also examine another important performance metric: the probability of exposing a failure with a small number of tests (specifically, ranging from 15 to 50). They found that \art has a significantly higher fault detection probability than \rand in this scenario. However, they note that the probability remains quite low, making it unlikely that a small test sample would reliably trigger a failure. Additionally, they prove a theorem showing that \art's advantage over \rand diminishes as the failure rate ($\theta$) increases, where diversity offers fewer benefits for fault detection.

Arcuri and Briand~\cite{ArcuriB11} consider also the possible advantages of \art over \rand when a single execution of the program under test is expensive. Assuming \art's capability of doubling the failure exposure probability $\theta$ (in line with the empirical evidence available from the literature), the cost of distance computation is amortized only when it is lower than half the cost of program execution. The condition is expressed as: 
$|W|\frac{1}{2\theta}((\frac{1}{2\theta} - 1) / 2)  t_d  + \frac{1}{2\theta}  t_e < \frac{1}{\theta}  t_e$, where $t_d$ represents the time for a single distance computation and $t_e$ the time for a single test case execution. In their example, with $\theta = 1.51 \cdot 10^{-5}$ and $|W| = 10$, this would require the test execution time to be 1.65 $\cdot$ 10$^5$ times longer than the distance computation time for \art to have an advantage, a scenario they consider highly unlikely in practice.

\section{Approach} \label{sec:approach}

\subsection{Framework}

In order to reduce the quadratic cost for distance computation required by the original \art algorithm, we need a faster way to compare each new candidate from $W$ (see \autoref{algo:art-dist}) with the archive of previously executed tests $Z$. The main idea behind our approach is that the archive $Z$ can be mapped to an aggregate representation $u$, which can be directly compared with each $w \in W$ without having to repeat the comparison for each $z \in Z$. The other assumption that we make is that the aggregate representation of the archive, $u$, can be efficiently updated whenever a new element $z$ is added to the archive $Z$. Without loss of generality, we make the assumption that $u$ belongs to ${\Bbb R}^k$, as commonly done with numeric embeddings of complex input spaces, with $k$ being the embedding dimensionality. Hence, we can formalize our framework as:

\begin{gather*}
  \exists g: {\Bbb P}(I) \rightarrow {\Bbb R}^k \quad \exists h: {\Bbb R}^k \times I \rightarrow {\Bbb R}^k \quad \exists d: {\Bbb R}^k \times I \rightarrow {\Bbb R} \\
  \quad \\
  \forall (u, i) \in {\Bbb R}^k \times I, u = g(Z) \\
  Z \in {\Bbb P}(I) \Rightarrow h(u, i) = g(Z \cup \{i\})
\end{gather*}

For a program under test $P$ that transforms an input $i \in I$ into an output $o \in O$, we assume we can define an aggregation function $g$ that transforms a set of input values from the powerset ${\Bbb P}(I)$ into its embedding form ${\Bbb R}^k$. This is the function that defines the aggregate representation $u$ of an archive of tests $Z$: $u = g(Z)$. The second assumption is that a new aggregate representation can be computed incrementally from a previous one through function $h$. Given the previous embedding $u$ of the current archive $Z$, for a new input $i \in I$ we can efficiently obtain the embedding of $Z \cup \{i\}$ by just applying $h$ to $u$ and $i$, i.e., by computing $h(u, i)$.
The last assumption of our framework is that a diversity function $d$ can be efficiently computed given the embedding $u \in {\Bbb R}^k$ of the archive $Z \in {\Bbb P}(I)$ and a new input $i \in I$.

There are several ways to instantiate our general framework, including incremental clustering, compression algorithms, or \grams. \grams are particularly appealing since they are conceptually simple and because they work on strings, and any program input can be represented as a string, thus increasing their applicability. Their simplicity also makes them easy to implement, which is crucial for ART, as it can allow higher performance. One possible embedding, $g$, based on \grams, is the \gram count. For example, consider an archive $Z$ containing the strings ``\texttt{aba}'', ``\texttt{abb}'', and ``\texttt{bc}''. The vocabulary of bigrams (with $Q = 2$) includes ``\texttt{ab}'', ``\texttt{ba}'', ``\texttt{bb}'', and ``\texttt{bc}''. The bigram count embedding of $Z$ would be: $G(Z) = \langle 2, 1, 1, 1 \rangle$, where the first entry reflects that ``\texttt{ab}'' appears twice, and the remaining bigrams appear once each.\footnote{In practice, this embedding is more efficiently represented as a dictionary rather than a vector, enabling also fast updates.}

The incremental embedding update function $h$ is simply a function that updates the bigram counts. For instance, given the embedding $G(Z) = \langle 2, 1, 1, 1 \rangle$, the input string ``\texttt{abc}'' determines the following update of the embedding: $\langle 2, 1, 1, 1 \rangle \rightarrow \langle 3, 1, 1, 2 \rangle$, as the two bigrams ``\texttt{ab}'' and ``\texttt{bc}'' appear once in the new input string.

Given the bigram counts of archive $Z$ and input $i$, there are various ways in which a diversity function $d$ can be defined, e.g., measuring  Gini impurity~\cite{BreimanFOS84} or  entropy~\cite{Shannon48} of the union bigram set. Let us consider entropy: given the embedding $G(Z) = \langle 2, 1, 1, 1 \rangle$ and the input string ``\texttt{abc}'', we can measure the entropy $H$ of the updated embedding $\langle 3, 1, 1, 2 \rangle$ by first converting counts into probabilities, and then applying the usual definition of entropy, i.e., $H(\langle 3/7, 1/7, 1/7, 2/7\rangle) = 1.84$.

\subsection{Algorithm}

\begin{algorithm}[ht]
\caption{\art with \gram aggregation} \label{algo:art-qgrams}

$Z$ = \{\} \;
\textit{add a random test case to $Z$ and execute it} \;
$\textit{Qcount}$ = \textit{computeQgramCounts}($Z$) \;
\While{\textit{stopping criterion not satisfied}}{
    \textit{sample set $W$ of random test cases} \;
    \ForEach{$w \in W$}{
        $w$.\textit{ent} = \textit{entropy}($\textit{Qcount}$.\textit{add}(\textit{computeQgramCounts}($w$))) \;
    }
    $w^*$ = $\arg\max_{w \in W}$ \{ \textit{$w$.ent} \} \;
    \textit{execute $w^*$ and add it to $Z$} \;
    $\textit{Qcount}$ = $\textit{Qcount}$.\textit{add}(\textit{computeQgramCounts}($w^*$)) \;
}

\end{algorithm}

\autoref{algo:art-qgrams} instantiates our framework with \grams. The aggregate representation $\textit{Qcount}$ of the previously executed inputs that are stored in the archive $Z$ is just the \gram counts, which can be efficiently represented as a dictionary mapping a \gram to its occurrence count (e.g., $\textit{Qcount}[\text{``}\texttt{ab}\text{''}] = 2$ if the bigram ``\texttt{ab}'' appears twice in $Z$). $\textit{Qcount}$ is initialized at Line~3 with the \gram counts of the initial, random input, added to $Z$ at Line~2.

Within the main test generation loop (Lines~4---12), the diversity of each new candidate $w$ from $W$ is determined by temporarily adding the \gram counts for $w$ to those previously computed for $Z$ and stored in the dictionary $\textit{Qcount}$, and then computing the entropy of the resulting counts (Line~7). The candidate input $w^*$ with maximum entropy, indicating higher diversity, is selected for execution (Lines~9--10) and its counts are used to update the dictionary of \gram counts (Line~11).

At each iteration of the main loop, \autoref{algo:art-qgrams} performs $|W|$ diversity (i.e., entropy) computations. If the algorithm stops after executing $\rho$ test cases, the number of diversity computations it requires is linear with $\rho$: $\sum_{i=1}^{\rho-1} |W| = |W|(\rho-1) = \Theta(\rho)$.

Let us consider the potential advantages of \art with \gram aggregation over \rand when the cost of executing a single test is high, following the analysis and example by Arcuri and Briand~\cite{ArcuriB11}. Assuming \art can double the failure exposure probability $\theta$, the cost of entropy computation now becomes justified when it is less than half the cost of a program execution:
$|W|(\frac{1}{2\theta} - 1) t_h + \frac{1}{2\theta} t_e < \frac{1}{\theta} t_e$,
where $t_h$ is the time for a single entropy computation, and $t_e$ is the time for a single test case execution. For $\theta = 1.51 \cdot 10^{-5}$ and $|W| = 10$, the test execution time would now need only to be 10 times longer than the entropy computation time--a scenario that is feasible in practice, e.g. when testing complex systems like multi-tier web applications.

\section{Simulation Results} \label{sec:sim-results}

We conducted our simulation experiments on a \texttt{Python} function that determines if the input string is palindrome, as our \grams based approach is directly applicable to string inputs. In particular, we chose $Q = 2$, measuring the diversity of the inputs using bigrams. We created a mutant of this function such that the failure probability (i.e., the probability of observing a different output in the original vs the mutant) depends on the maximum string length $L$. The failure probability for this mutant can be shown to be approximately $\theta = 1/L$. Hence, by choosing $L$ from the set $\{$100, 1,000, 10,000, 66,225 $\}$ we can explore the behaviour of \art with \gram aggregation (simply called \grams in the following) vs \rand when $\theta$ ranges from $10^{-2}$ to $1.51 \cdot 10^{-5}$, with the latter value matching exactly the failure probability considered by Arcuri and Briand in their simulation experiments~\cite{ArcuriB11}.

\begin{table}[ht]
	\centering
		
    \caption{Results of simulation experiments with no delay (left) and with 10 $ms$ delay (right). The failure probability is approximately \textit{1 / L}, with \textit{L} being the length of the input string. Results with relative standard error above the threshold ($0.05$) are in italics, while the highest P-measure and lowest F/T-measures values are bolded. Underlined values indicate statistical significance of the differences. The grey-colored rows separate the experiments with different lengths.}
    \footnotesize
    \label{table:simulation-results:simulation-results}
    \setlength{\tabcolsep}{2.7pt}
    \renewcommand{\arraystretch}{1.2}

    \begin{tabular}{rclrrrrr}
        \toprule

        \multicolumn{1}{l}{} 
        & \multicolumn{1}{l}{} 
        &  & \multicolumn{3}{c}{\textbf{No delay}} 
        & \multicolumn{2}{c}{\textbf{10 \textit{ms} delay}} \\
        
        \cmidrule(r){4-6}
        \cmidrule(r){7-8}

        \multicolumn{1}{c}{\textbf{L}} 
        & \multicolumn{1}{c}{\textbf{Fail. Prob.}}
        & \multicolumn{1}{c}{\textbf{Generator}} 
        & \multicolumn{1}{c}{\textbf{P-measure} $\uparrow$} 
        & \multicolumn{1}{c}{\textbf{F-measure} $\downarrow$} 
        & \multicolumn{1}{c}{\textbf{T-measure (s)} $\downarrow$} 
        & \multicolumn{1}{c}{\textbf{F-measure} $\downarrow$} 
        & \multicolumn{1}{c}{\textbf{T-measure (s)} $\downarrow$} \\

        \midrule
        
        100 & 1.00E-02 & \rand & 9.82E-03 & 100.2 & {\ul \textbf{0.00107}} & {\ul \textbf{98.2}} & {\ul \textbf{1.15}} \\
        100 & 1.00E-02 & \dist & {\ul \textbf{2.99E-02}} & {\ul \textit{\textbf{47.5}}} & \textit{5.566} & N/A & N/A \\
        100 & 1.00E-02 & \grams & 3.90E-04 & 122.9 & 0.1365 & 120.7 & 2.1 \\
        \rowcolor{mygrey} 1,000 & 1.00E-03 & \rand & 1.06E-03 & 985.24 & {\ul \textbf{0.1045}} & 985 & \textit{13.48} \\
        \rowcolor{mygrey} 1,000 & 1.00E-03 & \dist & {\ul \textbf{4.73E-03}} & \textit{252.16} & \textit{70.01} & N/A & N/A \\
        \rowcolor{mygrey} 1,000 & 1.00E-03 & \grams & 8.08E-04 & \textbf{167.93} & 0.4591 & {\ul \textbf{170.8}} & {\ul \textbf{3.33}} \\
        10,000 & 1.00E-04 & \rand & 1.09E-04 & 9,542.30 & {\ul \textit{\textbf{30.82}}} & \textit{9,461.70} & \textit{616} \\
        10,000 & 1.00E-04 & \grams & {\ul \textbf{2.63E-04}} & {\ul \textbf{1,050.50}} & \textit{76.83} & {\ul \textbf{\textit{1,200.20}}} & {\ul \textbf{\textit{40.16}}} \\
        \rowcolor{mygrey} 66,225 & 1.50E-05 & \rand & \textit{1.62E-05} & 63,815.60 & {\ul \textbf{3,103.2}} & \textit{58,397.70} & \textit{6,037.50} \\
        \rowcolor{mygrey} 66,225 & 1.50E-05 & \grams & \textit{\textbf{5.58E-05}} & {\ul \textit{\textbf{6,190.90}}} & \textit{5,259.80} & {\ul \textit{\textbf{4,826}}} & {\ul \textit{\textbf{2,520.8}}} \\

        \bottomrule
        
    \end{tabular}
\end{table}

Following the recommendations of Arcuri and Briand~\cite{ArcuriB11}, we consider three key performance metrics in our experiments: (1)~\textbf{P-measure}, the probability of exposing a fault (i.e., killing the mutant) within a limited number of test executions (50 in our case), where a higher value indicates better performance; (2)~\textbf{F-measure}, the number of test executions required to expose the fault, with a lower value being preferable; and (3)~\textbf{T-measure}, the total execution time, in seconds, before exposing the fault, where lower times are better.

While P-measure and F-measure gauge effectiveness without considering the time needed for diversity computation, T-measure is the decisive metric for comparing \art and \rand, as it accounts for the full computation time each algorithm requires to expose a fault. However, because the execution time for the palindrome-checking function in our setup is extremely low (on the order of $\mu s$), P-measure and F-measure still serve as useful indicators for scenarios where higher execution times might make ART's overhead for diversity computation worthwhile. To simulate such conditions, we introduced a 10 $ms$ delay into the palindrome-checking function, mimicking a long-running program. We report results for the function both with and without this delay.

We conducted most experiments on a MacBook Pro with an M3 chip and 24 GB of memory. For long-running experiments (e.g., those with $L = 66,225$), we used AWS, running them on an EC2 T3.large instance with 2 vCPUs, 8 GB of memory, and Ubuntu as the operating system. To ensure accurate time measurements, we used the same hardware for each value of $L$, allowing for consistent comparisons across failure probabilities. Rather than pre-defining the number of experiment repetitions, we incrementally increased repetitions until the relative standard error of the metrics (P-measure, F-measure, T-measure) fell below 0.05. In some cases, this required a substantial number of repetitions (up to 100,000). Overall, we completed 417,956 experiment repetitions, where each experiment involved either generating 50 test inputs (for P-measure) or generating inputs until a failure was observed (for F/T-measure). 

\autoref{table:simulation-results:simulation-results} shows the outcomes of our simulation experiments. For each maximum input string length (Column 1) \textit{L}, corresponding to a failure probability of approximately \textit{1 / L}, we compare \rand, \art using string edit distance (\dist), and \art with bigram aggregation (i.e., \grams) across the P-measure, F-measure, and T-measure metrics. When \textit{L} reaches 10,000 or 66,225, \dist becomes impractical due to the high cost of distance computation, so these rows exclude \dist. Values in italics indicate cases where the relative standard error could not be reduced below the 0.05 threshold within the experiment’s 3-month time budget. For each failure probability, the highest P-measure and lowest F/T-measure values are bolded, and underlined when statistically significant differences, determined by the Wilcoxon rank sum test~\cite{wilcoxon1945individual} at $\alpha = 0.05$, are observed between methods.

The probability of exposing a fault (P-measure) is generally higher for \dist compared to \rand and \grams, but \dist does not scale beyond \textit{L =} 1,000. At higher failure probabilities (\textit{L =} 100), \grams not only perform poorly but actually hinder performance, with a P-measure significantly lower than that of \rand. Since \dist performs well in this scenario, one possible explanation could be that bigrams are simply  less effective at capturing diversity for small inputs. As failure probabilities decrease, however, \grams become increasingly effective, outperforming other techniques by $2.6 \times$ at \textit{L =} 10,000 and $3.4 \times$ at \textit{L =} 66,225. In such cases, random approaches struggle to hit the small failing region, while diversity helps the generator better explore and cover it.

The F-measure results align with the ones for the P-measure. At high failure probabilities (low \textit{L}), \dist requires the fewest executions to expose a fault, followed by \rand and then \grams. However, as failure probabilities decrease, \dist becomes impractical, and the diversity introduced by \grams proves beneficial. At \textit{L =} 10,000 and \textit{L =} 66,225, \grams accelerates fault discovery by roughly $10 \times$ compared to \rand. This demonstrates that \grams scales well to lower failure probabilities, requiring fewer executions than \rand to uncover faults.

The T-measure metric indicates that \rand is the fastest, followed by \grams and \dist, which exceeds the former by a huge amount, becoming inapplicable at \textit{L =} 10,000. Since results in \autoref{table:simulation-results:simulation-results} indicate that \rand has a decreasing speedup over \grams as the failure probability is reduced, we conjecture that at low failure probabilities \grams might become convenient when the execution time of the unit under test is substantially higher than the execution of a single \texttt{Python} function (on average, the palindrome function requires 1.8 $\mu s$ to execute). We simulated this by introducing a delay of 10 $ms$ (last two columns of \autoref{table:simulation-results:simulation-results}), making the execution $5k \, \times$ higher than without delay. While at high failure probabilities (\textit{L =} 100) \rand remains convenient, we observe a substantial advantage of \grams already at \textit{L =} 1,000, where \grams is $4 \times$ faster.

\begin{tcolorbox}[boxrule=0pt,frame hidden,sharp corners,enhanced,borderline north={1pt}{0pt}{black},borderline south={1pt}{0pt}{black},boxsep=2pt,left=2pt,right=2pt,top=2.5pt,bottom=2pt]
	\textbf{Summary}: Simulation experiments demonstrate that the \art with \gram aggregation algorithm scales effectively to low failure probabilities, exposing faults with a higher probability and requiring fewer test executions than random test generation. However, the linear cost of \gram computation is only offset, making \grams the fastest in terms of clock time, when the unit under test has a non-negligible execution time (e.g., 10 ms).
\end{tcolorbox}

\section{Experimental Results} \label{sec:results}

\subsection{Subjects}

\begin{table}[ht]
	\centering

    \footnotesize
    \caption{Web application subjects implemented with different JS Frameworks~\cite{Biagiola0RT19}.}
    \label{table:results:subjects}
    \setlength{\tabcolsep}{3.1pt}
    \renewcommand{\arraystretch}{1.2}

    \begin{tabular}{llrrr}
        \toprule

        \multicolumn{1}{c}{\textbf{Subject}} 
        & \multicolumn{1}{c}{\textbf{JS Framework}} 
        & \multicolumn{1}{c}{\textbf{LOC (JS)}} 
        & \multicolumn{1}{c}{\textbf{Stars}} 
        & \multicolumn{1}{c}{\textbf{Commits}} \\

        \midrule
        
        retroboard & React & 2,144 & 775 & 966 \\
        dimeshift & Backbone & 5,140 & 194 & 204 \\
        phoenix & React & 2,289 & \textgreater{}2,500 & 422 \\
        splittypie & Ember.js & 2,710 & 174 & 350 \\
        petclinic & AngularJS & 2,939 & 200 & 295 \\
        pagekit & Vue.js & 4,214 & \textgreater{}5,500 & 4,933 \\
        
        \bottomrule
        
    \end{tabular}
\end{table}

\autoref{table:results:subjects} shows the six subject web applications we considered for our empirical study. We considered all subjects in the benchmark proposed by Biagiola et al.~\cite{Biagiola0RT19}, which have also been used in other studies on web application testing~\cite{krishna2023fragment,zheng2021automatic}.
Such applications have been implemented with different Javascript (JS) frameworks, and their size is representative of modern web applications that use a JS framework~\cite{ocariza2017detecting}.

\subsection{Test Generation for Web Applications}

The web testing approach proposed by Biagiola et al.~\cite{Biagiola0RT19} use \textit{Page Objects} (POs) to model web pages using classes. Each PO encapsulates all the methods that are responsible for the interaction with the web page it represents, in an End-to-End (E2E) manner, i.e., through its Graphical User Interface. The PO pattern was introduced by Fowler~\cite{folwer2013page} to reduce the maintainability effort when the web application evolves~\cite{leotta2013improving,leotta2020family}, and the Selenium web testing framework strongly encourages its use when developing E2E web tests~\cite{selenium2024page}. From POs it is possible to extract a navigation model of the web application, where nodes are POs and edges are the methods that bring the application from a web page modeled with one PO to another~\cite{biagiola2017search}. 

\begin{figure*}[ht]
    \centering

    \resizebox{0.75\textwidth}{!}{
    \includegraphics[scale=0.37]{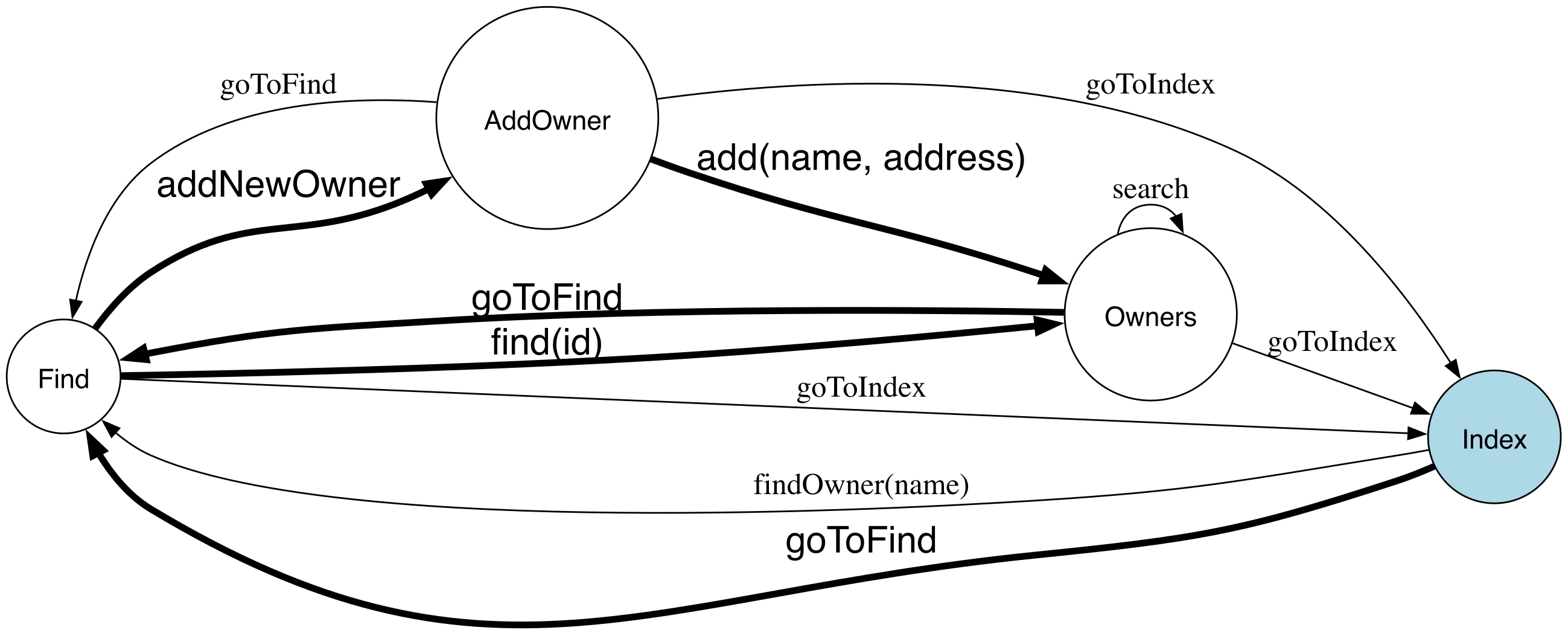}}
    
    \caption{Portion of the navigation model of the \texttt{petclinic} web application. Nodes represent pages of the web application, while edges represent the actions that bring the application from one page to another. The home page of the web application is highlighted in blue, while the path highlighted in bold represents an example of a \textit{feasible} navigation path.} 
    \label{fig:results:graph-petclinic} 
\end{figure*}

\autoref{fig:results:graph-petclinic} shows a portion of the navigation model of the \texttt{petclinic} application, one of the subjects we used in our study. The home page is modeled by the \textit{Index} PO, highlighted in blue, with two methods, i.e., \texttt{goToFind}, with no input arguments, and \texttt{findOwner}, that has one input argument namely the \texttt{name} of the pet owner to look for. The other two methods with input arguments in the model are the \texttt{find} method of the \texttt{Find} PO, and the \texttt{add} method of the \texttt{AddOwner} PO, respectively with arguments \texttt{id} and \texttt{add}, \texttt{address}. 
The path \texttt{goToFind} $\rightarrow$ \texttt{addNewOwner} $\rightarrow$ \texttt{add} $\rightarrow$ \texttt{goToFind} $\rightarrow$ \texttt{find}, highlighted in bold, is an example of a \textit{feasible} navigation path. Indeed, the methods in the navigation graph may be constrained by one or more \textit{guards}, also called preconditions. A guard is a boolean condition that depends on the state of the web application, and on the input values of the method. The state of the web application includes the DOM, and any data that is stored persistently during the interaction. Biagiola et al.~\cite{Biagiola0RT19} define a path as feasible iff there exists a set of inputs for all the methods in the path such that these methods can be executed correctly, i.e., all the guards are satisfied. For instance, the highlighted path  is feasible as it is possible to satisfy all its guards, while the path \texttt{goToFind} $\rightarrow$ \texttt{find} is not feasible, as no owner was added during the execution, and there exists no \texttt{id} value that can satisfy the precondition of the \texttt{find} method of the \texttt{Find} PO.

Given a web application modeled with POs and its navigation model, a test generator extracts abstract navigation paths by traversing the model and makes them concrete by generating input values for the methods involved. For instance, a test case for the highlighted path in \autoref{fig:results:graph-petclinic} is $t = \langle \texttt{goToFind}, \texttt{addNewOwner}, \texttt{add}(\text{``}\texttt{John}\text{''}, \text{``}\texttt{My street}\text{''})$ $,\texttt{goToFind}, \texttt{find(0)} \rangle$. Hence, the challenge for a test generator in this context is that of extracting feasible paths and generating concrete tests with the appropriate input values. 

Biagiola et al.~\cite{Biagiola0RT19} use the \art framework described in \autoref{algo:art-dist} with sequence edit distance as a distance function to address the test generation problem. They considered two versions of their diversity based test generator, where the distance function is either the method sequence edit distance, or a combination of method sequence edit distance and input value distance. In this paper, we only describe and experiment with the sequence-only version (called \dist in the following), as their results show that the two are equivalent. For instance, let us consider three test cases, i.e., $t_1 = \langle \texttt{goToFind}, \texttt{goToIndex} \rangle$, $w_1 = \langle \texttt{findOwner}(\text{``}\texttt{John}\text{''}), \texttt{goToIndex} \rangle$, and $w_2 = \langle \texttt{goToFind}, \texttt{find}(\texttt{1}), \texttt{goToFind}, \texttt{find}(\texttt{2}) \rangle$, where $t_1$ is already executed (i.e., $t_1 \in Z$ following the notation of \autoref{algo:art-dist} and \autoref{algo:art-qgrams}), while $w_1$ and $w_2$ are two candidate test cases (i.e., $w_1, \, w_2 \in W$). To choose which one to execute, we evaluate the sequence edit distance w.r.t. $t_1$,  obtaining a distance of 1 for $w_1$, as there is one non-matching action in $t_1$ to be replaced with an action from $w_1$ (i.e., \texttt{goToFind} in $t_1$, replaced by \texttt{findOwner} in $w_1$), and a distance of 3 for $w_3$. Hence, \dist selects $w_2$ as the next test to execute, as it is the farthest from $t_1$.

We considered two versions of \grams, i.e., sequence-only (\gramss for short), and sequence plus inputs (\gramssi for short). In this example and in the experiments, we considered $Q = 2$, i.e., bigrams. First, we compute the \gram count for $t_1$, shown in \autoref{table:results:qgram-counts}, which is the same for both versions, as the two methods in the bigram (``\texttt{goToFind}'', ``\texttt{goToIndex}'') have no input arguments.

\begin{table}[ht]
	\centering
		
    \caption{\gram count of \gramss (left) and \gramssi (right) for the running example. Each row represents a bigram ($Q = 2$), i.e., a tuple of two method strings, while the columns show the counts of the corresponding bigram in the respective test. The last row shows the entropy  $H$ for the two candidates $w_1$ and $w_2$. \matteo{fix the spacing}}
    \label{table:results:qgram-counts}
    \footnotesize
    \setlength{\tabcolsep}{3.1pt}
    \renewcommand{\arraystretch}{1.2}
    
    \begin{subtable}[t]{0.49\textwidth}
    
        \centering
        \begin{tabular}{llcc}
            \toprule
    
            & \multicolumn{3}{c}{\gramss} \\

            \cmidrule{2-4}
            
            & \multicolumn{1}{c}{$t_1$} & $w_1$ & $w_2$ \\
    
            \midrule
            
            (``\texttt{goToFind}'', ``\texttt{goToIndex}'') & \multicolumn{1}{c}{1} & 1 & 1 \\
            (``\texttt{findOwner}'', ``\texttt{goToIndex}'') &  & 1 & \multicolumn{1}{l}{} \\
            (``\texttt{goToFind}'', ``\texttt{find}'') &  & \multicolumn{1}{l}{} & 2 \\
            (``\texttt{find}'', ``\texttt{goToFind}'') &  & \multicolumn{1}{l}{} & 1 \\
    
            \midrule
    
            \multicolumn{1}{r}{\textit{H}} &  & \textit{1.0} & \textit{1.5} \\
            
            \bottomrule
            
        \end{tabular}
    
    \end{subtable}
    \hfill
    \begin{subtable}[t]{0.49\textwidth}

        \centering
        \begin{tabular}{llcc}
            \toprule
    
            & \multicolumn{3}{c}{\gramssi} \\

            \cmidrule{2-4}
            
            & \multicolumn{1}{c}{$t_1$} & $w_1$ & $w_2$ \\
    
            \midrule
            
            (``\texttt{goToFind}'', ``\texttt{goToIndex}'') & \multicolumn{1}{c}{1} & 1 & 1 \\
            (``\texttt{findOwner(John)}'', ``\texttt{goToIndex}'') &  & 1 & \multicolumn{1}{l}{} \\
            (``\texttt{goToFind}'', ``\texttt{find(1)}'') &  & \multicolumn{1}{l}{} & 1 \\
            (``\texttt{find(1)}'', ``\texttt{goToFind}'') &  & \multicolumn{1}{l}{} & 1 \\
            (``\texttt{goToFind}'', ``\texttt{find(2)}'') &  & \multicolumn{1}{l}{} & 1 \\
    
            \midrule
    
            \multicolumn{1}{r}{\textit{H}} &  & \textit{1.0} & \textit{2.0} \\
            
            \bottomrule
            
        \end{tabular}

    \end{subtable}
\end{table}

Concerning \gramss, the \gram count for $w_1$ adds the bigram (``\texttt{findOwner}'', ``\texttt{goToIndex}''), while $w_2$ adds two more bigrams (second and third row of the leftmost table) w.r.t. $t_1$, of which the bigram (``\texttt{goToFind}'', ``\texttt{find}'') appears twice. In contrast, \gramssi treats this as two distinct bigrams because the input values to the \texttt{find} method differ at different points in the sequence ($\texttt{id} = 1$ in the second position, and $\texttt{id} = 2$ in the fourth position). In both cases, the approach selects candidate $w_2$ for execution, as its entropy is higher compared to $w_1$. 

\subsection{Research Questions and Metrics}

The \textit{goal} of our empirical study is to evaluate the test generation capabilities of \art with \gram aggregation (\grams) in comparison to two competing techniques: \art with method sequence edit distance (\dist) and random testing (\rand).

\textbf{RQ\textsubscript{1} [Coverage]} \textit{How does \grams compare to \dist and \rand in terms of coverage of the target web applications, under a fixed test generation time budget?}

In this research question, we explore whether real-world web applications possess the characteristics that make \grams more effective than its competitors when given a fixed test generation budget (8 hours in our experiments).

\head{Metrics} The coverage metric we use is the number of edges covered in the navigation graph of the web application under test. This measures a technique's ability to generate test scenarios that traverse deeply nested and hard-to-reach web pages. 

\textbf{RQ\textsubscript{2} [Efficiency]} \textit{How does \grams compare to \dist and \rand in terms of the speed at which coverage grows as the number of generated tests increases?}

With this research question, we investigate how quickly coverage expands, helping us assess whether greater diversity leads to faster coverage of test targets. 

\head{Metrics} We measure the Area Under the Curve (AUC) from the plot that relates the number of executed tests to the coverage achieved. Additionally, we calculate  AUC@20\%, which represents the area under the curve when only 20\% of the tests are executed. This accounts for scenarios with limited test generation budgets. We use a percentage instead of an absolute number of tests to accommodate variations in test execution times across different subjects.

\textbf{RQ\textsubscript{3} [Uniqueness]} \textit{To what extent does \grams cover unique test targets that the competing test generation techniques fail to cover?}

This research question explores \grams' ability to reach test targets that other approaches might miss. Since these unique targets could potentially contain faults, covering them suggests an increased likelihood of exposing bugs that other techniques might overlook. 

\head{Metrics} For each test generation repetition, we measure the number of test targets that are uniquely covered by each technique.

\textbf{RQ\textsubscript{4} [Sequence Length]} \textit{What is the relationship between the length of the test sequence and its selection by \grams or \dist?}

RQ\textsubscript{4} examines whether any of the considered \art techniques achieve high diversity simply by generating longer test sequences, which may not necessarily cover new test targets. 

\head{Metrics} We measure the length of the test sequences selected by \grams or \dist over the test generation runs.

\subsection{Experimental Procedure}

In our empirical evaluation, we used the navigation graphs provided by Biagiola et al.~\cite{Biagiola0RT19} for the respective web applications, as well as their implemented POs. Given a navigation model for a certain web application, we follow the original paper and traverse the graph with a random walk of a length chosen randomly in the interval $[1, L]$, where $L = 40$ is the maximum length. At each iteration, we randomly select an edge/method in the graph that is not yet covered, we generate a random walk starting from the index PO, and we connect the last node of the random walk with the target node of the uncovered edge. Considering \dist, \gramss and \gramssi, we generate at each iteration $|W| = 30$ candidates, and we execute the one that increases diversity the most according to the different methods (i.e., method sequence edit distance for \dist and entropy for \gramss and \gramssi). For \gramss and \gramssi, we use $Q = 2$, i.e., we keep track of bigram counts during the execution. 

We compared four techniques in our empirical study, i.e., \rand, \dist, \gramss, and \gramssi, executing each of them with a time budget of eight hours. To account for the inherent randomness of the generation process, we executed each technique five times, and compared the results using rigorous statistical tests~\cite{arcuri2014hitchiker}. In particular, we used the Wilcoxon rank sum test~\cite{wilcoxon1945individual} to compute the $p$-value, with the confidence threshold $\alpha = 0.05$, and the Vargha-Delaney effect size ($\hat{A}_{12}$)~\cite{vargha-delaney} to assess the magnitude of the difference. Overall, we have six web applications, four techniques, an eight hour time budget, and five repetitions for a total CPU time of $6 \times 4 \times 8h \times 5 = 960h$ (or 40 CPU days).

\subsection{Results}

\begin{table}[ht]
	\centering
		
    \caption{Results for RQ\textsubscript{1}, RQ\textsubscript{2}, and RQ\textsubscript{3}. Web applications are ranked according to their \textit{complexity}, shown in the first two columns. The grey-colored rows represent medium-complexity web applications, separating low-complexity (above), and high-complexity (below) applications. Bolded values represent the best average for each metric, while underlined values represent a $p$-value $< 0.05$, and at least a medium effect size magnitude, of \grams w.r.t. \rand and \dist. \new{For each of the four evaluation metrics (AUC, AUC@20\%, Unique Targets, Coverage) and for each web app, bolded values represent the best average among \texttt{Rand}, \texttt{Dist}, \gramssi, \gramss. Sometimes the best average is a tie involving 2 or more techniques (e.g., \texttt{pagekit} has a tie on AUC@20, with the best average value of 0.14 reached by \textsc{Dist}, \gramssi, and \gramss).}}
    \footnotesize
    \label{table:results:rq1-rq2-rq3}
    \setlength{\tabcolsep}{2.7pt}
    \renewcommand{\arraystretch}{1.2}

    \begin{tabular}{rcccccccccccccccccccc}
        \toprule

        \multicolumn{1}{l}{} 
        & \multicolumn{2}{c}{\textbf{Cpx.}} 
        & \multicolumn{4}{c}{\textbf{\rand}} 
        & \multicolumn{4}{c}{\textbf{\dist}} 
        & \multicolumn{4}{c}{\textbf{\gramssi}}
        & \multicolumn{4}{c}{\textbf{\gramss}} \\
        
        \cmidrule(r){2-3}
        \cmidrule(r){4-7}
        \cmidrule(r){8-11}
        \cmidrule(r){12-15}
        \cmidrule(r){16-19}

        \multicolumn{1}{l}{} 
        & \rot{\textbf{\# Targets}}
        & \rot{\textbf{\# Diff. Targets}}
        & \rot{\textbf{Coverage (\%)}}
        & \rot{\textbf{AUC}}
        & \rot{\textbf{AUC@20\%}}
        & \rot{\textbf{Unique Targets}}
        & \rot{\textbf{Coverage (\%)}} 
        & \rot{\textbf{AUC}} 
        & \rot{\textbf{AUC@20\%}} 
        & \rot{\textbf{Unique Targets}} 
        & \rot{\textbf{Coverage (\%)}} 
        & \rot{\textbf{AUC}} 
        & \rot{\textbf{AUC@20\%}} 
        & \rot{\textbf{Unique Targets}} 
        & \rot{\textbf{Coverage (\%)}} 
        & \rot{\textbf{AUC}} 
        & \rot{\textbf{AUC@20\%}} 
        & \rot{\textbf{Unique Targets}} \\

        \midrule
        
        retroboard & 29 & 0 & \textbf{93.10} & 0.84 & 0.14 & \textbf{0.60} & 91.03 & \textbf{0.88} & \textbf{0.16} & 0.40 & 87.59 & 0.84 & \textbf{0.16} & 0.20 & 89.66 & 0.85 & 0.15 & 0.40 \\
        dimeshift & 72 & 0 & \textbf{88.61} & 0.81 & 0.13 & 1.00 & 87.78 & \textbf{0.85} & \textbf{0.14} & 0.80 & 86.94 & 0.81 & \textbf{0.14} & \textbf{1.40} & 84.72 & 0.75 & 0.12 & 1.20 \\
        \rowcolor{mygrey} phoenix & 38 & 5 & 75.79 & 0.71 & 0.13 & 0.20 & 73.16 & 0.72 & 0.13 & 0.00 & 80.00 & {\ul 0.77} & \textbf{0.14} & 0.60 & {\ul \textbf{90.00}} & {\ul \textbf{0.84}} & {\ul \textbf{0.14}} & {\ul \textbf{3.80}} \\
        \rowcolor{mygrey} splittypie & 44 & 6 & 82.73 & 0.77 & 0.12 & 0.20 & 82.27 & 0.80 & 0.14 & 0.20 & {\ul \textbf{91.36}} & {\ul \textbf{0.87}} & \textbf{0.15} & {\ul \textbf{1.60}} & 87.73 & 0.82 & 0.14 & 0.20 \\
        \rowcolor{mygrey} petclinic & 47 & 9 & 80.00 & 0.75 & 0.13 & 0.00 & 80.00 & 0.79 & 0.15 & 0.60 & \textbf{83.40} & \textbf{0.82} & \textbf{0.16} & \textbf{1.20} & 82.55 & \textbf{0.82} & \textbf{0.16} & 1.00 \\
        pagekit & 212 & 34 & 80.28 & 0.76 & 0.13 & 0.40 & 78.30 & 0.76 & \textbf{0.14} & 0.80 & {\ul 81.89} & {\ul 0.79} & \textbf{0.14} & 1.40 & {\ul \textbf{83.59}} & {\ul \textbf{0.80}} & \textbf{0.14} & {\ul \textbf{3.60}} \\

        \midrule

        \textit{Avg} & \textit{---} & \textit{---} & \textit{83.42} & \textit{0.77} & \textit{0.13} & \textit{0.40} & \textit{82.09} & \textit{0.80} & \textit{0.14} & \textit{0.47} & \textit{85.20} & \textit{0.82} & \textit{0.15} & \textit{1.07} & \textit{86.38} & \textit{0.81} & \textit{0.14} & \textit{1.70} \\

        \bottomrule
        
    \end{tabular}
\end{table}

\head{RQ\textsubscript{1} [Coverage]} The first two columns of \autoref{table:results:rq1-rq2-rq3} display the \textit{complexity} of each web application subject. We define complexity based on the number of \textit{difficult} coverage targets (Column~2) and, as a tiebreaker, the total number of coverage targets (Column~1), which corresponds to the number of transitions in the navigation graphs. Difficult targets are those that \rand fails to cover in any of the five repetitions. For example, \texttt{retroboard} has 29 coverage targets, all of which are covered by \rand in at least one repetition. In contrast, \texttt{pagekit} has 212 coverage targets, with \rand failing to cover 34 of them across all repetitions. Web applications are ranked by complexity in \autoref{table:results:rq1-rq2-rq3}, with \texttt{retroboard} and \texttt{dimeshift} being the least complex, \texttt{phoenix}, \texttt{splittypie}, and \texttt{petclinic} having intermediate complexity, and \texttt{pagekit} as the most complex. This setup mirrors the simulation experiments in \autoref{sec:sim-results}, where instead of faults with varying failure probabilities, we here deal with coverage targets that have varying probabilities of being reached.

Columns~3, 7, 11, and 15 of \autoref{table:results:rq1-rq2-rq3}, show the average coverage, as a percentage of the total number of targets, achieved in five repetitions by \rand, \dist, \gramssi and \gramss respectively, for each web application subject. We observe that in the web application subjects with low complexity, i.e., \texttt{retroboard}, and \texttt{dimeshift}, \rand has the highest coverage on average, although the difference with the other techniques is not statistically significant. On the other hand, for intermediate and high complexity web applications, \grams techniques outperform both \rand and \dist in most cases. Indeed, for \texttt{phoenix}, \gramss covers significantly more than \rand and \dist with a large effect size (90\% vs 75.79\% and 73.16\% respectively), while the difference with \gramssi is not statistically significant (\gramssi with an average coverage of 80\% is significantly different from \dist with a large effect size). For \texttt{splittypie}, \gramssi covers significantly more than \rand and \dist with a large effect size (91.36\% vs 82.73\% and 82.27\% respectively), while the difference with \gramss (87.73\%) is not statistically significant. For \texttt{petclinic}, \gramssi has the highest coverage, being very close to that of \gramss (i.e., 83.40\% vs 82.55\%), although the difference with \rand and \dist is not statistically significant (both have an average coverage of 80\%). For \texttt{pagekit}, both \gramssi and \gramssi  cover significantly more than \rand and \dist, with a large effect size (81.89\% and 83.59\% vs 80.28\% and 78.30\%); although \gramss has the highest average coverage, the difference with \gramssi is not statistically significant. 

\begin{figure}[ht]
    \begin{subfigure}{0.49\textwidth}
        \centering
            \includegraphics[scale=0.068]{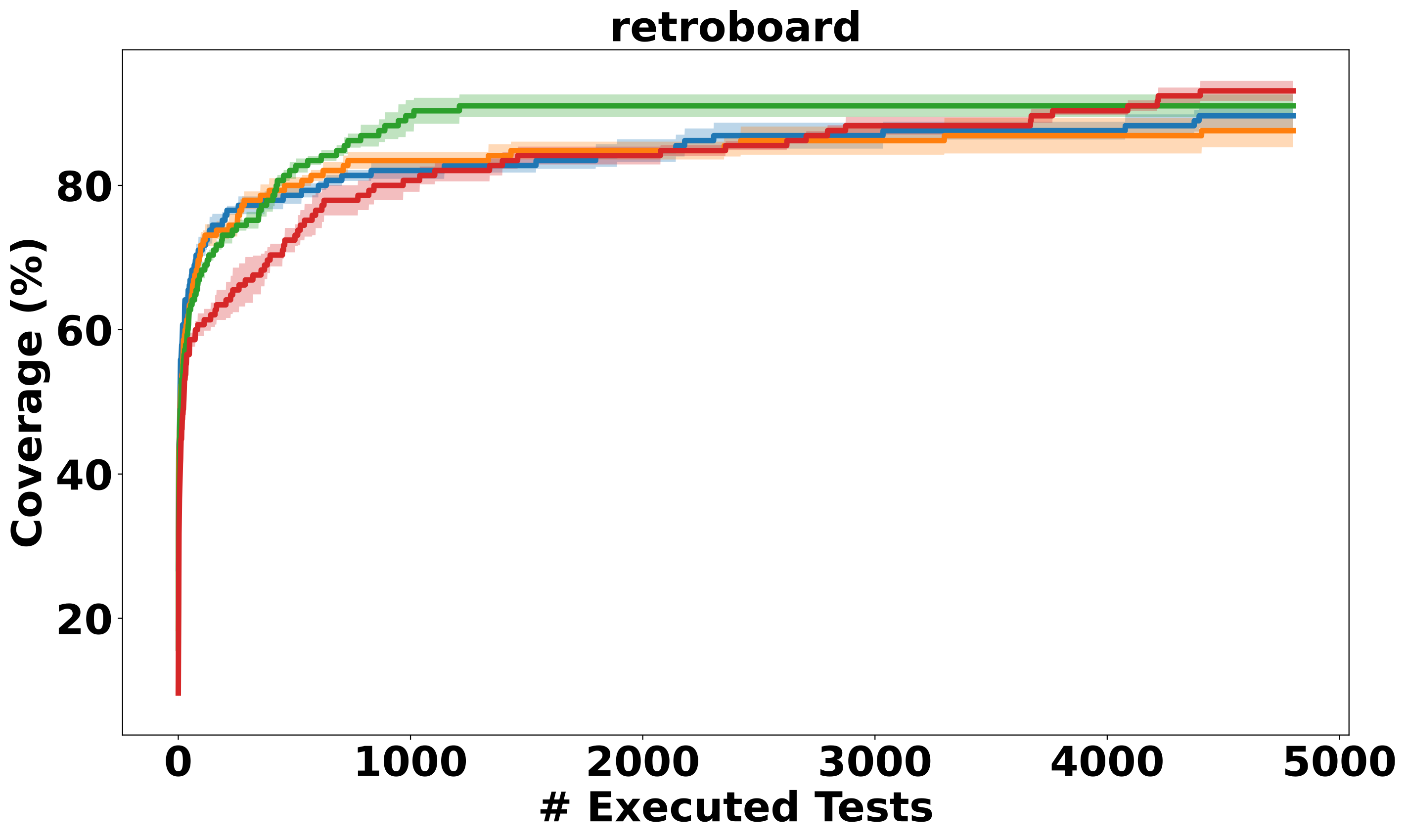}
    \end{subfigure}
    \begin{subfigure}{0.49\textwidth}
        \centering
            \includegraphics[scale=0.068]{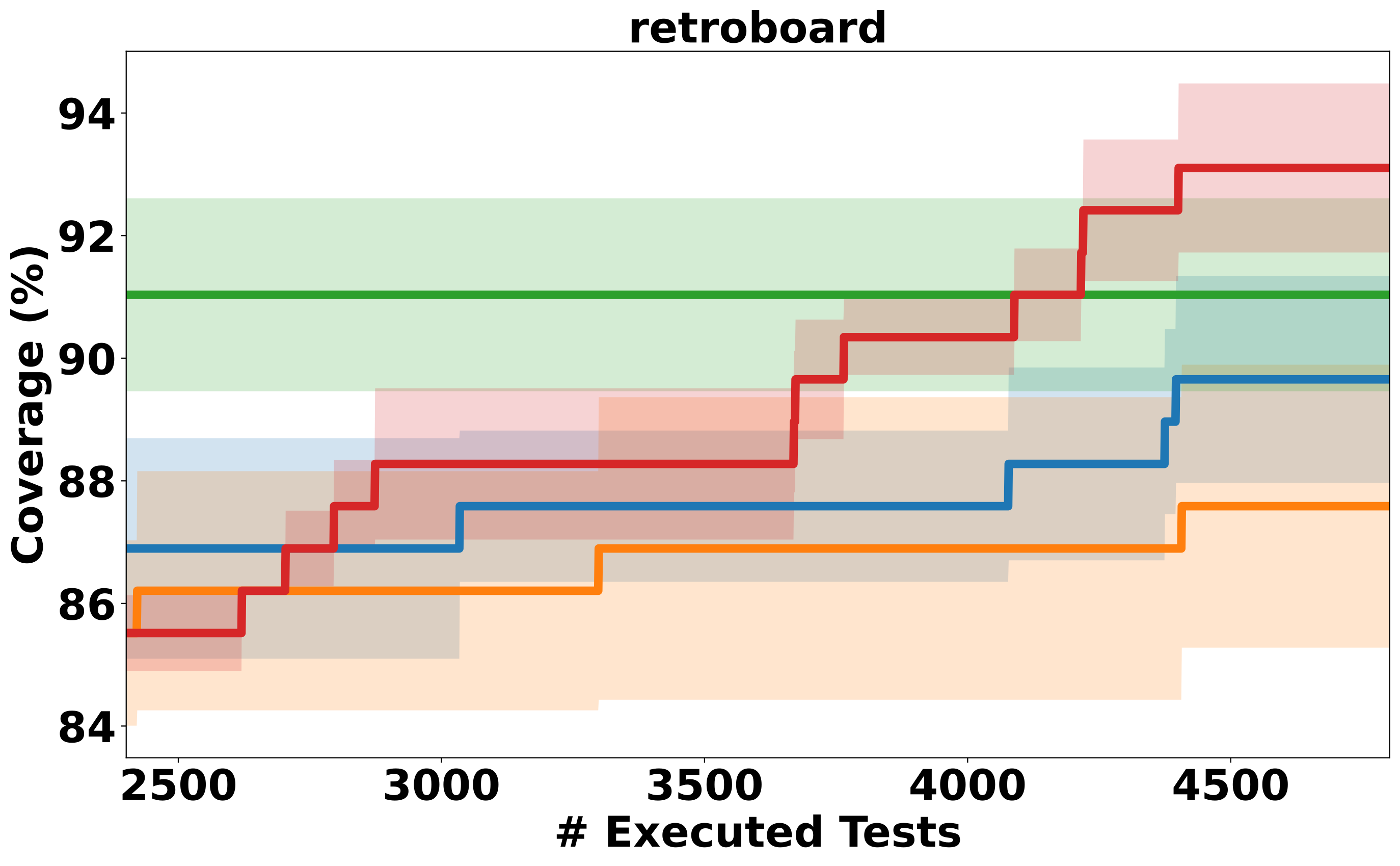}
    \end{subfigure}
    \\
    \begin{subfigure}{0.49\textwidth}
        \centering
            \includegraphics[scale=0.068]{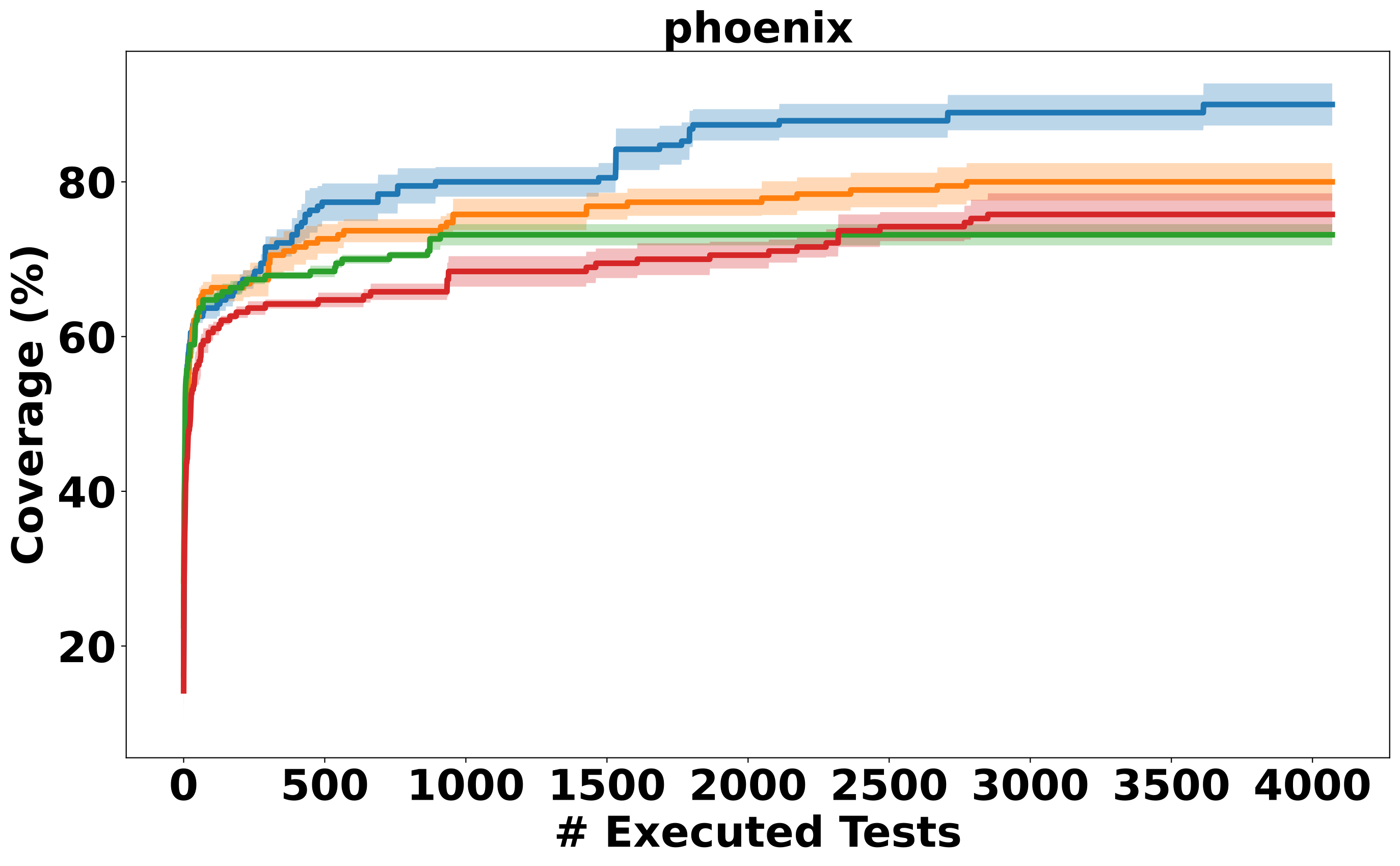}
    \end{subfigure}
    \begin{subfigure}{0.49\textwidth}
        \centering
            \includegraphics[scale=0.068]{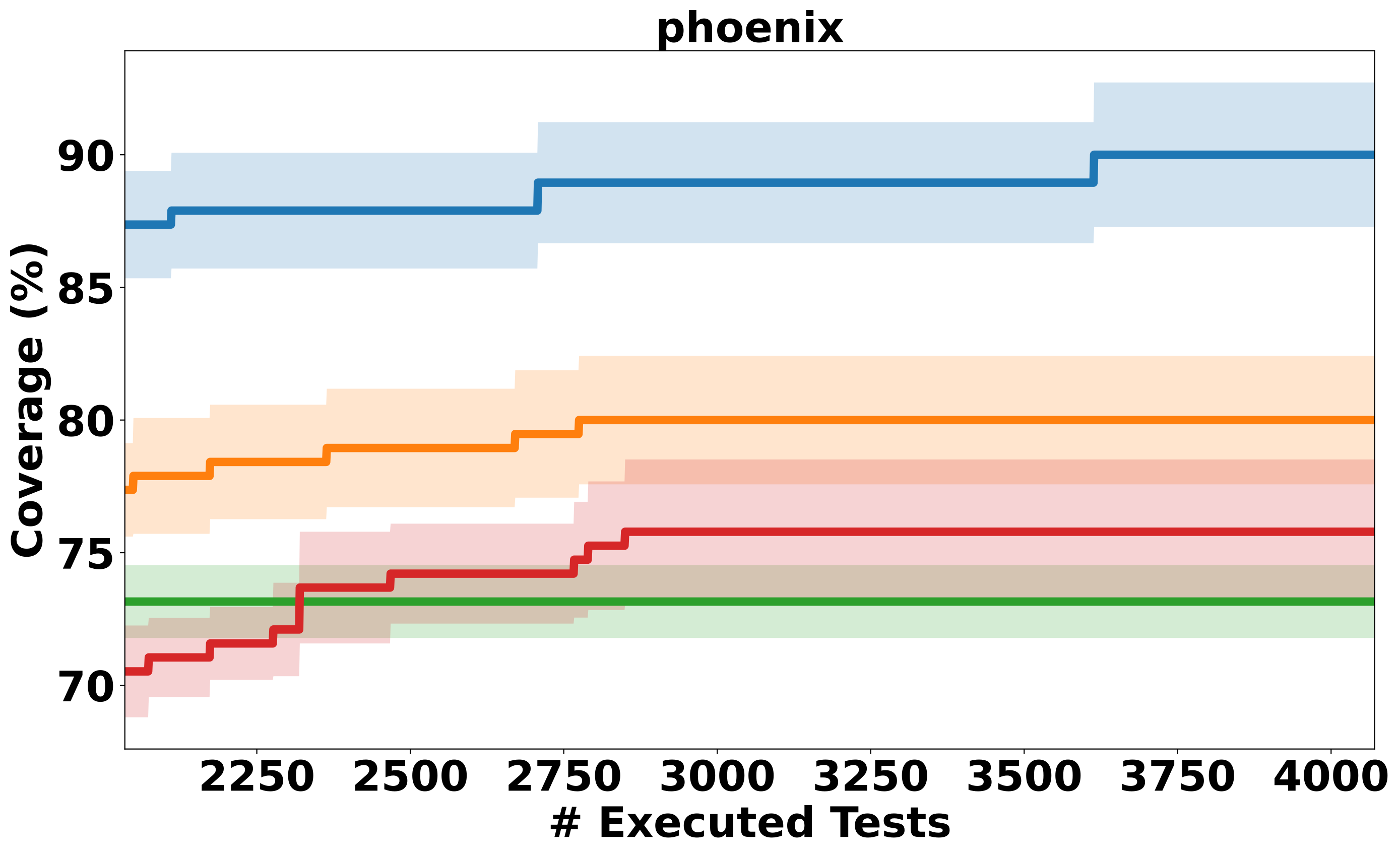}
    \end{subfigure}
    \\
    \begin{subfigure}{0.49\textwidth}
        \centering
            \includegraphics[scale=0.068]{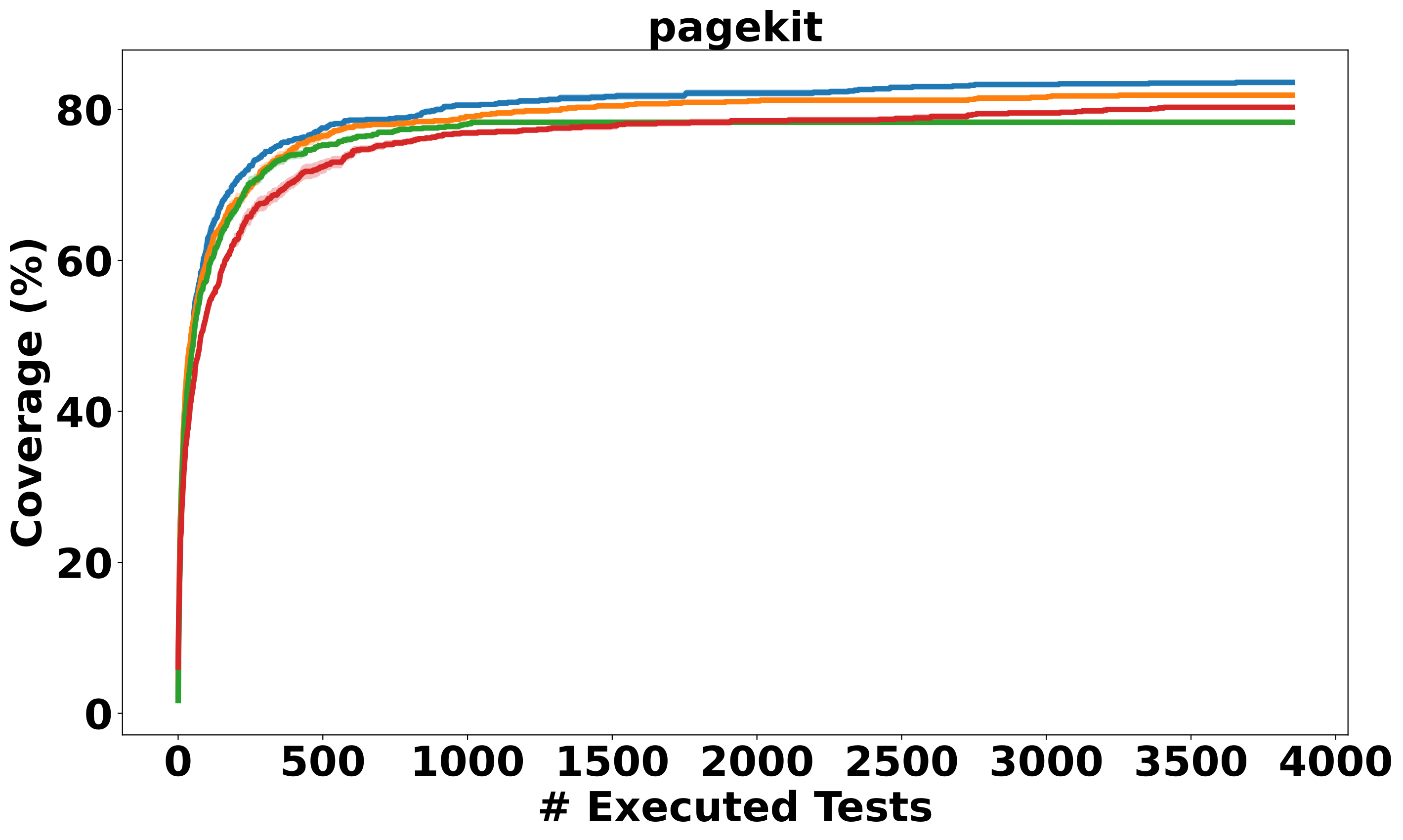}
    \end{subfigure}
    \begin{subfigure}{0.49\textwidth}
        \centering
            \includegraphics[scale=0.068]{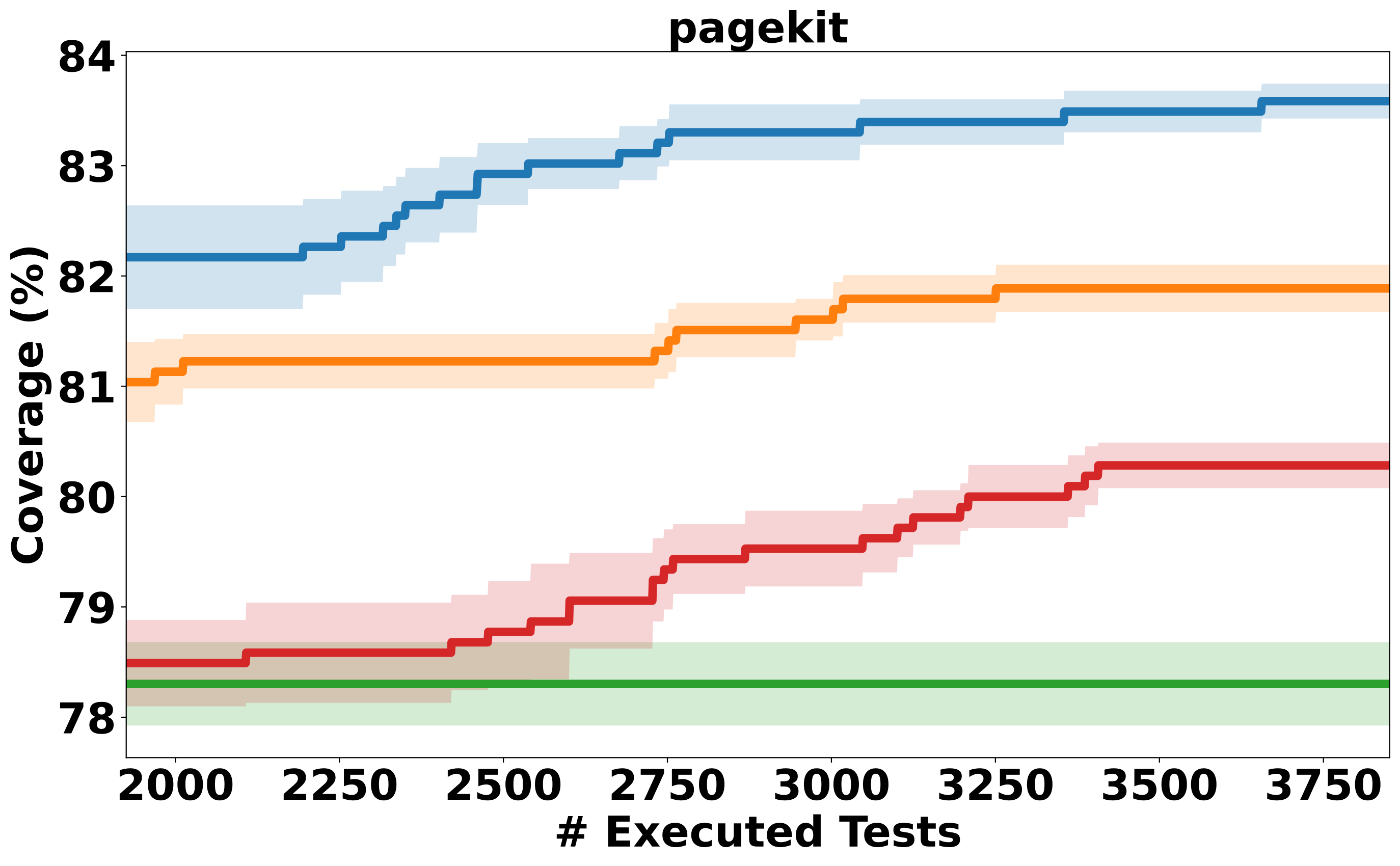}
    \end{subfigure}

    \caption{Coverage over number of executed tests for three subjects and all techniques \colorline{myred} \rand, \colorline{mygreen} \dist, \colorline{myblue} \gramss, \colorline{myorange} \gramssi. Solid lines represent the average over five repetitions, while the shaded areas around them represent the standard error of the mean. Curves are padded with the respective last values, such that all techniques have curves with the same number of points. The left-hand side shows the coverage trend considering all executed test cases, with a zoomed-in view of the latest half of the executed test cases on the right. Best viewed in color.} 
    \label{fig:results:coverage-over-time} 
\end{figure}

\autoref{fig:results:coverage-over-time} illustrates the coverage trends over the total number of executed tests. The solid lines represent the average coverage across five repetitions, while the shaded areas indicate the standard error of the mean. To ensure all techniques are displayed on the same plot, we aligned the data to the technique that executed the highest number of tests (i.e., the longest sequence). Shorter sequences were padded with their corresponding final values to match the length of the longest sequence.

Due to space constraints, we selected three representative web applications of varying complexity: \texttt{retroboard} (low complexity), \texttt{phoenix} (intermediate complexity), and \texttt{pagekit} (high complexity). Full results are available in the replication package~\cite{replication-package}. For each web application, the full plot is shown on the left, and a zoomed-in view of the last half of the test executions is on the right.

In \texttt{retroboard}, all techniques converge closely by the end, as shown by the overlapping uncertainties, with no technique clearly outperforming the others, though \rand (red line) achieves the highest average coverage. In \texttt{phoenix}, which contains harder-to-reach targets for \rand, the techniques are more spread out, with \gramss (blue line) further apart, and \gramssi (orange line) slightly overlapping with \rand. In the high-complexity subject \texttt{pagekit}, the separation between \grams and the other techniques becomes even more pronounced, with both \gramss and \gramssi showing a clear advantage over \rand and \dist, with no overlap.

\begin{tcolorbox}[boxrule=0pt,frame hidden,sharp corners,enhanced,borderline north={1pt}{0pt}{black},borderline south={1pt}{0pt}{black},boxsep=2pt,left=2pt,right=2pt,top=2.5pt,bottom=2pt]
	\textbf{RQ\textsubscript{1} [Coverage]}: In intermediate and high complexity web applications, \grams techniques outperform both \rand and \dist. In three out of four such cases, this improvement is statistically significant, with coverage gains ranging from 4\% (\texttt{petclinic}) to 18\% (\texttt{splittypie}).
\end{tcolorbox}

\head{RQ\textsubscript{2} [Efficiency]} Columns~4, 8, 12, and 16 of \autoref{table:results:rq1-rq2-rq3}, show the average area under the curve (AUC) over five repetitions for all the techniques and subjects. Areas have been normalized between 0 and 1, by dividing them by the maximum area (the area of the rectangle with base the maximum number of executed tests and height 100). The trend that we observe is similar to the coverage results. In low complexity web application subjects, \dist has the highest AUC (0.88 and 0.85 in \texttt{retroboard} and \texttt{dimeshift} respectively); \dist is significantly better than \rand (0.84) and \gramss (0.85) with a large effect size in \texttt{retroboard}, and it is significantly better than \gramss (0.75) with a large effect size in \texttt{dimeshift}. In \texttt{phoenix} and \texttt{pagekit} both \gramssi and \gramss are significantly better than \rand and \dist with a large effect size. \gramss has the highest average in both subjects (i.e., 0.84 and 0.80 respectively); the difference with \gramssi (0.77 and 0.79 respectively) is statistically significant only in \texttt{pagekit}. In \texttt{splittypie}, \gramssi (0.87) is significantly better than \rand (0.77), \dist (0.80) and \gramss (0.82), while in \texttt{petclinic} both \gramssi and \gramss achieve the best average AUC (i.e., 0.82), although only the difference with \rand (0.75) is statistically significant with a large effect size (the difference with \dist, with an AUC of 0.79, is not).

Columns~5, 9, 13, and 17 of \autoref{table:results:rq1-rq2-rq3}, show the average area under the curve when considering 20\% of the executed tests (AUC@20\%). In this context, the differences are less pronounced than when considering AUC (only in \texttt{phoenix}, \gramss is significantly better than \rand and \dist), although all diversity techniques significantly outperform \rand in five out of six applications (except \texttt{dimeshift}, and in \texttt{pagekit} \dist is not statistically different from \rand). Interestingly, \gramssi seems to always have the best AUC@20\% average, that in low complexity subjects is matched by \dist, while in intermediate to high complexity is matched by \gramss (except in \texttt{splittypie}, where \gramssi has the best average of all).

On the left hand side of \autoref{fig:results:coverage-over-time}, we have a visual appreciation of the efficiency results. In particular, at the beginning of each plot (metric AUC@20\%), diversity-based techniques quickly cover more targets (i.e., they are more efficient) than \rand.

\begin{table}[ht]
	\centering
		
    \caption{Test generation statistics. From left to right, the table shows average values of number of executed tests (\# Exec. Tests), length of a test case measured as number of statements (Length (\# Stmts)), executed time for each test in seconds (Exec. Time (s)), and time to generate a test in seconds (Gen. Time (s)).}
    \footnotesize
    \label{table:results:other-metrics}
    \setlength{\tabcolsep}{2.9pt}
    \renewcommand{\arraystretch}{1.2}

    \begin{tabular}{rcccccccccccccccc}
        \toprule

        \multicolumn{1}{l}{} 
        & \multicolumn{4}{c}{\textbf{\rand}} 
        & \multicolumn{4}{c}{\textbf{\dist}} 
        & \multicolumn{4}{c}{\textbf{\gramssi}}
        & \multicolumn{4}{c}{\textbf{\gramss}} \\
        
        \cmidrule(r){2-5}
        \cmidrule(r){6-9}
        \cmidrule(r){10-13}
        \cmidrule(r){14-17}

        \multicolumn{1}{l}{} 
        & \rot{\textbf{\# Exec. Tests}} 
        & \rot{\textbf{Length (\# Stmts)}} 
        & \rot{\textbf{Exec. Time (s)}} 
        & \rot{\textbf{Gen. Time (s)}} 
        & \rot{\textbf{\# Exec. Tests}} 
        & \rot{\textbf{Length (\# Stmts)}} 
        & \rot{\textbf{Exec. Time (s)}} 
        & \rot{\textbf{Gen. Time (s)}} 
        & \rot{\textbf{\# Exec. Tests}} 
        & \rot{\textbf{Length (\# Stmts)}} 
        & \rot{\textbf{Exec. Time (s)}} 
        & \rot{\textbf{Gen. Time (s)}} 
        & \rot{\textbf{\# Exec. Tests}} 
        & \rot{\textbf{Length (\# Stmts)}} 
        & \rot{\textbf{Exec. Time (s)}} 
        & \rot{\textbf{Gen. Time (s)}} \\

        \midrule
        
        retroboard & 4,584.6 & 33.04 & 2.99 & 0.01 & 1,260.0 & 59.13 & 3.45 & 16.28 & 4,618.0 & 15.81 & 2.49 & 0.35 & 4,750.2 & 14.76 & 2.41 & 0.32 \\
        dimeshift & 4,136.8 & 34.69 & 3.57 & 0.01 & 1,192.0 & 58.80 & 3.98 & 17.06 & 4,047.2 & 32.98 & 3.36 & 0.38 & 3,962.0 & 30.34 & 3.56 & 0.33 \\
        phoenix & 3,339.0 & 55.59 & 4.93 & 0.03 & 1,089.4 & 88.09 & 6.08 & 17.12 & 3,571.0 & 41.92 & 3.73 & 0.92 & 3,686.4 & 42.73 & 3.70 & 0.84 \\
        splittypie & 4,618.8 & 36.28 & 2.81 & 0.01 & 1,257.8 & 60.70 & 3.02 & 16.76 & 4,247.0 & 31.91 & 2.91 & 0.49 & 4,468.2 & 36.12 & 2.82 & 0.40 \\
        petclinic & 2,929.6 & 39.14 & 6.28 & 0.02 & 1,038.6 & 63.65 & 8.42 & 16.10 & 3,608.0 & 43.71 & 4.10 & 0.50 & 2,566.6 & 58.84 & 6.89 & 0.44 \\
        pagekit & 3,654.8 & 35.51 & 4.53 & 0.03 & 1,135.0 & 57.98 & 5.29 & 17.02 & 3,274.8 & 42.15 & 4.63 & 0.81 & 3,409.2 & 38.51 & 4.41 & 0.74 \\

        \midrule

        \textit{Avg} & \textit{3,877.3} & \textit{39.04} & \textit{4.19} & \textit{0.02} & \textit{1,162.1} & \textit{64.73} & \textit{5.04} & \textit{16.72} & \textit{3,894.3} & \textit{34.75} & \textit{3.54} & \textit{0.58} & \textit{3,807.1} & \textit{36.88} & \textit{3.97} & \textit{0.51} \\

        \bottomrule
        
    \end{tabular}
\end{table}

\autoref{table:results:other-metrics} shows four statistics related to the test generation process for each technique and each subject. Particularly useful to interpret the efficiency results are the number of executed tests (Columns~1, 5, 9, and 13) and the time to generate a test case (Columns~4, 8, 12, and 16). The generation time for diversity-based techniques consists of the time to generate the $|W| = 30$ candidate test cases (see \autoref{algo:art-dist} and \autoref{algo:art-qgrams}), and the diversity maximization computation to select the test case that will be executed. On the other hand, the generation time for \rand only takes into account the time to generate a single test case. On average, considering all web application subjects (last row of \autoref{table:results:other-metrics}), \rand only spends 0.02 $s$ in generating the test to execute, while \dist is much more expensive, taking on average 16.7 $s$ to generate a test case, as it scales quadratically with the number of executed test cases. On the other hand, \grams techniques are very efficient when selecting a test case to execute, taking on average 0.58 $s$ and 0.51 $s$ for \gramssi and \gramss respectively, as they scale linearly with the number of executed tests. The cost for generating and selecting a test reflects on the total number of executed tests; indeed, \dist executes much less tests than \rand on average (i.e., \textit{1,162} vs \textit{3,877}), while \grams techniques execute a comparable number of tests (\textit{3,894} and \textit{3,807} respectively for \gramssi and \gramss). 

The different number of  tests executed by each technique is visible in \autoref{fig:results:coverage-over-time}, where right-padding indicates that the budget finished (the curve was extended for the purpose of comparison with the other techniques). \dist spends most of the budget for diversity computation, with its coverage curves flattening much earlier than the competing techniques. On the other hand, \rand, by executing many more tests than \dist, is able to achieve a higher overall coverage than \dist if given enough time, although it is less efficient when a low test generation budget is available. \grams techniques combine the best of both worlds, i.e., the efficiency of \dist at low budget regimes (AUC@20\%), and the ability to execute a comparable number of tests as \rand, at high budget regimes. This combination allows \grams techniques to be more efficient overall (AUC), especially in intermediate and high complexity subjects.

\begin{tcolorbox}[boxrule=0pt,frame hidden,sharp corners,enhanced,borderline north={1pt}{0pt}{black},borderline south={1pt}{0pt}{black},boxsep=2pt,left=2pt,right=2pt,top=2.5pt,bottom=2pt]
	\textbf{RQ\textsubscript{2} [Efficiency]}: \grams techniques are significantly more efficient than \rand in five out of six subjects when the test generation budget is low (AUC@20\%). With a higher test generation budget, \grams techniques outperform both \rand and \dist in intermediate and high complexity subjects, achieving statistical significance in three out of four cases, with AUC improvements ranging from 3.7\% (\texttt{petclinic}) to 16.7\% (\texttt{phoenix}).
\end{tcolorbox}

\head{RQ\textsubscript{3} [Uniqueness]} Columns~6, 10, 14 and 18 of \autoref{table:results:rq1-rq2-rq3}, show the average number of coverage targets that a certain technique uniquely covers in each repetition. For instance, \rand covers on average 0.60 coverage targets in \texttt{retroboard}, that \dist, \gramssi and \gramss do not cover; \rand has the highest average of uniquely covered targets in \texttt{retroboard}, but the difference with the other techniques is not statistically significant. In \texttt{dimeshift}, \gramssi has the highest average (1.40), above \gramss (1.20), \rand (1.0) and \dist (0.80); again the difference with the other techniques is not significant. In \texttt{phoenix}, \gramss  covers significantly more unique targets (3.80) than all the other techniques with a large effect size, with \gramssi ranking second (0.60). Similarly, \gramss covers significantly more unique targets (3.60) than the competing techniques  with a large effect size  in \texttt{pagekit}, with again \gramssi ranking second (1.40). On the other hand, \gramssi performs better than \gramss in \texttt{splittypie} and \texttt{petlicnic} (1.60 and 1.20 vs 0.20 and 1.0 respectively); the difference with the other techniques is significant only in \texttt{splittypie}.

\begin{tcolorbox}[boxrule=0pt,frame hidden,sharp corners,enhanced,borderline north={1pt}{0pt}{black},borderline south={1pt}{0pt}{black},boxsep=2pt,left=2pt,right=2pt,top=2.5pt,bottom=2pt]
	\textbf{RQ\textsubscript{3} [Uniqueness]}: \grams techniques outperform \rand and \dist in intermediate and high complexity web applications, achieving statistical significance in three out of four cases. In the worst case  (\texttt{pagekit}), \grams techniques cover $4.5\times$ more unique targets than the best of \rand and \dist, while in the best case (\texttt{phoenix}), they cover $19\times$ more unique targets.  
\end{tcolorbox}

\head{RQ\textsubscript{4} [Sequence Length]} Columns~2, 6, 10, and 14 of \autoref{table:results:other-metrics} show the average length (measured as number of statements) of the test cases selected for execution by each technique. We can observe that \dist selects the longest test cases, with an average length of 64.73 statements. On the other hand, \rand, \gramss and \gramssi are quite comparable, featuring an average length of 39.04, 34.75, and 36.88 statements respectively. The length of the selected test case correlates with the time it takes to execute it, shown in Columns~3, 7, 11, and 15 of \autoref{table:results:other-metrics}. \dist has the highest average with 5.04 $s$, followed by \rand with 4.19 $s$, \gramss with 3.97 $s$ and \gramssi with 3.54 $s$. The length of the selected test case is the reason why \grams techniques are able to execute a comparable number of tests w.r.t. \rand, despite the higher generation time. In fact, the tests selected by \grams techniques are overall shorter than the tests generated by \rand (the only exception being \texttt{petclinic}, where both \gramssi and \gramss select longer tests, and \texttt{pagekit}, where both select slightly longer tests), compensating the longer generation time. 

\begin{figure}[ht]
    \begin{subfigure}{0.49\textwidth}
        \centering
            \includegraphics[scale=0.068]{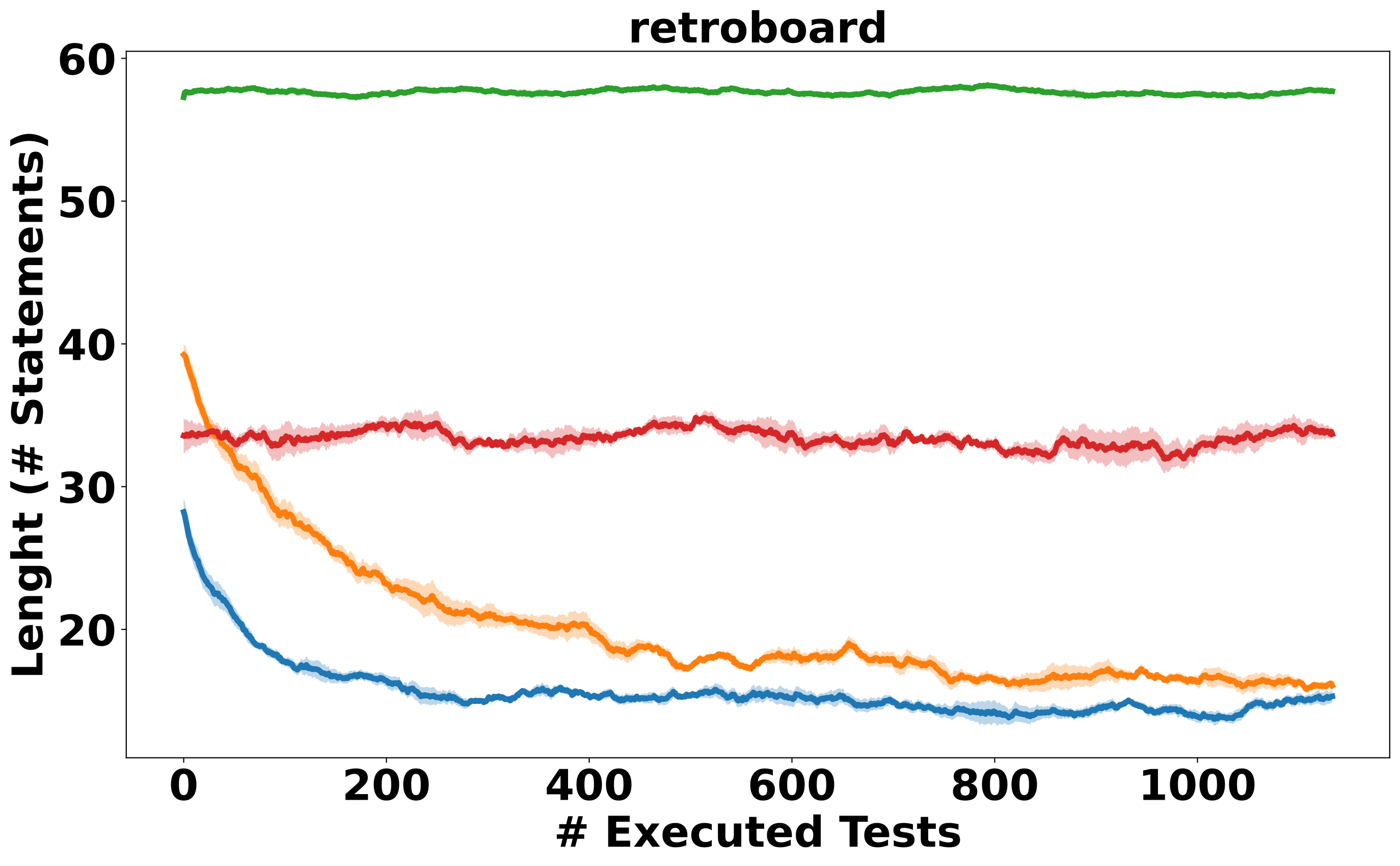}
    \end{subfigure}
    \begin{subfigure}{0.49\textwidth}
        \centering
            \includegraphics[scale=0.068]{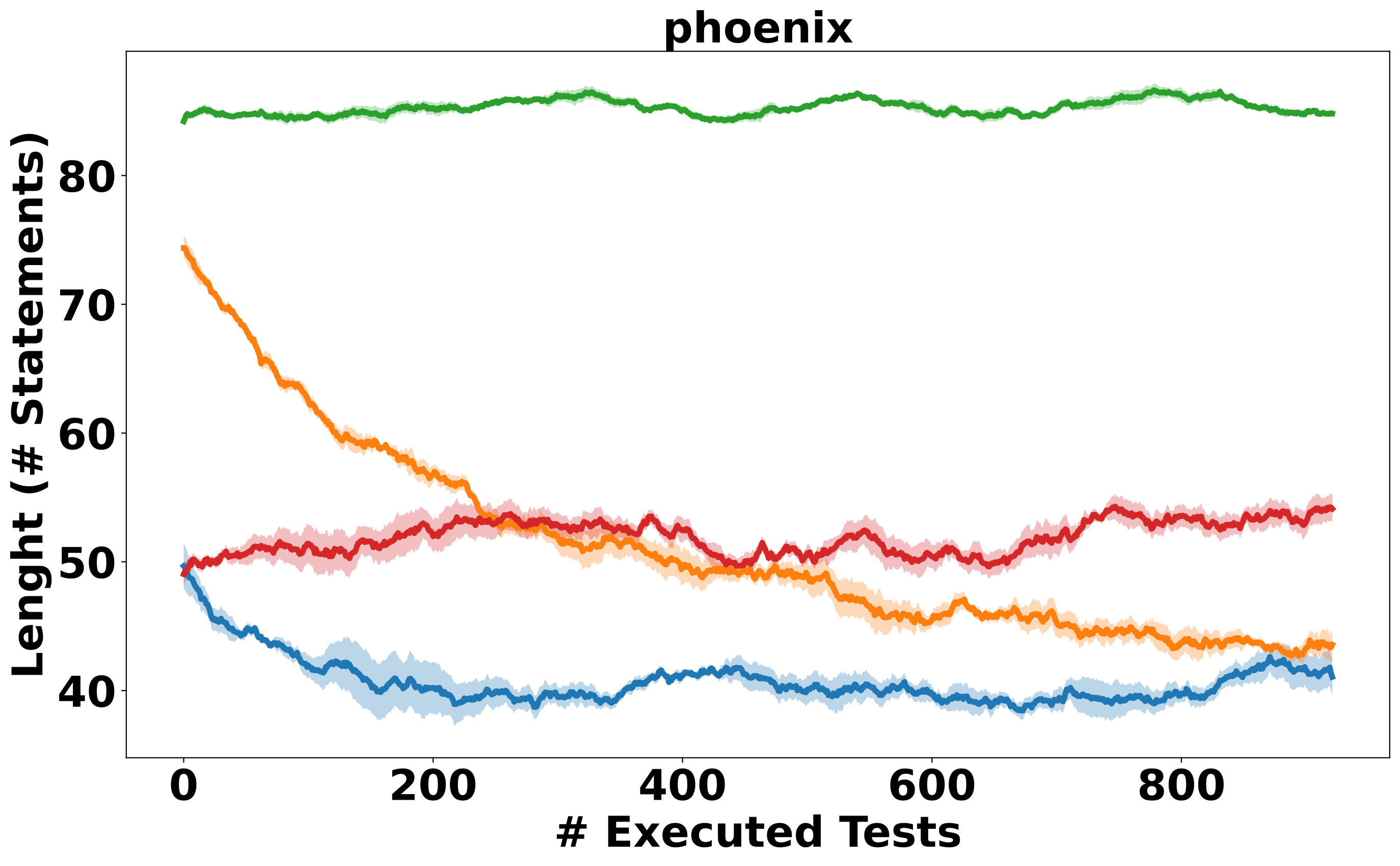}
    \end{subfigure}
    \\
    \begin{subfigure}{\textwidth}
        \centering
            \includegraphics[scale=0.068]{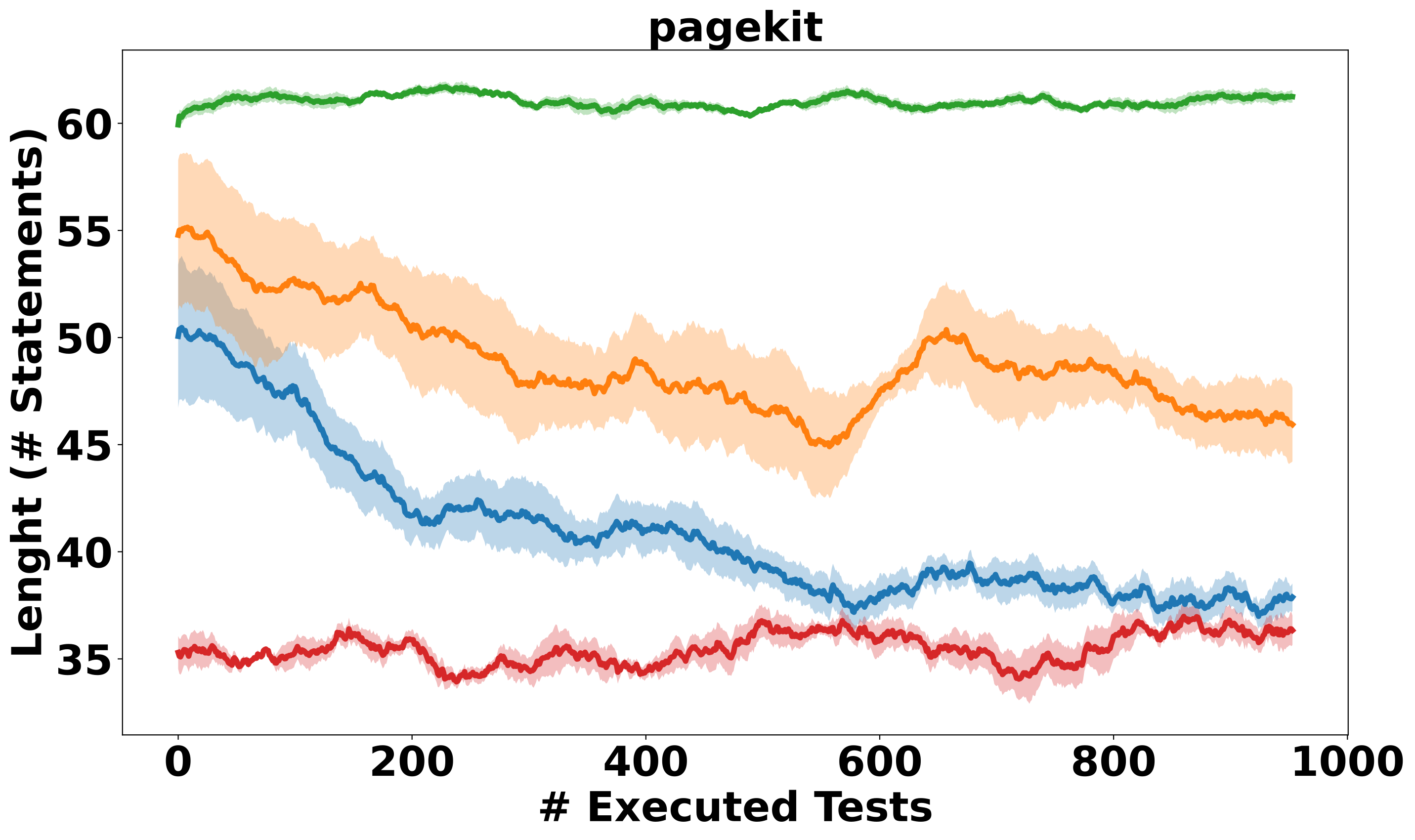}
    \end{subfigure}

    \caption{Length of the selected tests over the number of executed tests for three subjects and all techniques \colorline{myred} \rand, \colorline{mygreen} \dist, \colorline{myblue} \gramss, \colorline{myorange} \gramssi. The length (y-axis) is expressed as number of statements. Solid lines represent the average over five repetitions, while the shaded areas around them represent the standard error of the mean. Each point is smoothed with a window size of 100, such that the trend is more visible. Curves are truncated at the minimum number of executed tests across techniques, to display the same number of points for all techniques. Best viewed in color.} 
    \label{fig:results:length-over-time} 
\end{figure}

\autoref{fig:results:length-over-time} shows the trends (smoothed over a window of size 100) of the selected test case length over the number of executed tests of each technique for three subjects of different complexity, i.e., \texttt{retroboard}, \texttt{phoenix} and \texttt{pagekit} (full results are available in the replication package~\cite{replication-package}). In all the plots, we observe that \dist tends to select the longest test cases to maximize diversity, and this does not change as the test generation progresses. This is natural given that a longer sequence allows more room for a high edit distance. The length of the test cases generated by \rand is lower, but it also remains roughly constant as the number of executed tests increases. On the other hand, both \grams techniques tend to select long tests at the beginning of the generation process, while the length decreases as executed tests accumulate. Indeed, maximization of  the sequence edit distance can be achieved by selecting long test cases, while maximizing the entropy of the \grams requires selecting unexplored, but not necessarily long, sequences of statements. While selecting long sequences is beneficial to \grams at the beginning of the generation process, to maximize diversity and to quickly cover high probability targets, hard to cover targets need a more focused selection, which is facilitated by shorter, unexplored, sequences.

\begin{tcolorbox}[boxrule=0pt,frame hidden,sharp corners,enhanced,borderline north={1pt}{0pt}{black},borderline south={1pt}{0pt}{black},boxsep=2pt,left=2pt,right=2pt,top=2.5pt,bottom=2pt]
	\textbf{RQ\textsubscript{4} [Sequence Length]}: \dist selects long test cases from the beginning to the end of the test generation process. On the other hand, \grams techniques tend to select long tests at the beginning, while privileging shorter tests at the end, focusing the generation budget towards unexplored sequences.
\end{tcolorbox}

\subsection{Discussion}

Through theoretical analysis, simulations, and experiments on six web applications, we have demonstrated that \gram aggregation can transform the effectiveness of \art from an ``illusion'' into a practical reality. While random testing or traditional \art can be effective at high failure probabilities or for less complex programs under test, the \art with \gram aggregation techniques we have proposed have benefits for finding faults with lower failure probabilities or when programs under test are of medium to high complexity. 

\head{\gramss vs \gramssi} Our results show that considering input values leads to a slightly higher efficiency, especially in low budget scenarios (i.e., AUC@20\%). This might be explained by the fact that \gramssi tend to select slightly longer tests than \gramss throughout the generation process (see \autoref{fig:results:length-over-time}). On the other hand, considering method sequences only, seems to be more effective coverage-wise. Indeed, in the cases when \gramssi is better, the difference is small (3.6\% in \texttt{splittypie}, and 0.85\% in \texttt{petclinic}), while there is a larger difference when \gramss is better (1.7\% in \texttt{pagekit}, where the difference is statistically significant, and 10\% in \texttt{phoenix}). Moreover, \gramss tends to cover more unique targets than \gramssi on average (i.e., 1.70 vs 1.07 considering all the subjects). Although more experiments are needed to understand the impact of input values on such metrics, our results suggest that \gramss is the preferred \grams technique, unless efficiency, especially at low budget regimes, is a priority.

\head{Number of Candidates for Diversity Computation} In our experiments, we set the number of candidates to $|W| = 30$, keeping it fixed throughout the test generation process. While a smaller candidate pool may be sufficient early on for covering high-probability targets--making \art with \gram aggregation resemble \rand--a larger pool might prove beneficial later in the process for selecting test cases that reach harder-to-cover targets. Dynamically adjusting the number of candidates could improve overall coverage, a topic that future work should explore further.

\head{\new{High-dimensional Input Spaces}} \new{\grams can be applied to arbitrary high-dimensional input spaces, as an input in such space can be represented as a sequence of values that can be broken down into sub-sequences of length q. For instance, let us consider a function \textit{f(int a, string b, float c)}, with inputs $a=25$, $b=\texttt{``xxy''}$, $c=12.3$. The inputs $(a, b, c)$ can be represented as the sequence $\langle B_i, \texttt{``x''}, \texttt{``x''}, \texttt{``y''}, B_j\rangle$, where numerical inputs $a=25$ and $c=12.3$ are mapped into pre-defined bins $B_i$, $B_j$, and the string input is broken down into single characters. The bins are organized into a sequence, for example $B_1$, \dots, $B_k$, where each bin corresponds to a specific range or interval in the numerical space. If $q=2$, we would have $(B_i, \texttt{``x''})$ as the first \gram, $(\texttt{``x''}, \texttt{``x''})$ as the second \gram, and so on. }

\head{\new{\grams Limitations}} \new{While \grams automate the manual effort required for ad-hoc partitioning of the input space, they only focus on syntactic differences. A semantically meaningful partitioning would require manual, human knowledge~\cite{zhang2016prioritization}. Moreover, we acknowledge that usage of \grams with complex data types might have limitations, which would require further investigation, particularly because \grams are limited to local properties of the input sequence.} 

\subsection{Threats to validity}

\head{External validity} We evaluated our approach using a benchmark of six web applications and with specific \gram choices, which raises questions about generalization. However, our empirical findings align with the theoretical analysis, suggesting that generalization may be feasible for scenarios that meet the theoretical assumptions behind our approach--primarily the ability to define \grams that allow the aggregation to be updated incrementally.

\head{Conclusion validity} Following the best practices in empirical software engineering, we support our conclusions by executing the experiments multiple times, and by assessing the statistical significance of the differences using statistical tests and effect size measures~\cite{arcuri2014hitchiker}.

\section{Conclusion and Future Work} \label{sec:conclusion}

We propose a novel framework for Adaptive Random Testing (\art) that measures diversity through an aggregation function updated incrementally. We instantiated this framework using \gram counts as the aggregation function and entropy as the diversity measure. Our theoretical analysis shows that \art with \grams reduces the algorithmic cost from quadratic to linear, making the cost of diversity computation easily offset when test execution times are sufficiently long. Simulation results with a 10 $ms$ test execution time, along with empirical results from six real-world web applications, support these findings, demonstrating that \art with \gram aggregation is more effective and efficient than both \art with distance computation and random generation, particularly for web applications with hard-to-cover targets.

In future work, we plan to explore alternative instantiations of our framework, such as using other diversity metrics like Gini impurity~\cite{rokach2005top} or compression based approaches~\cite{feldt2016tsdm}. We also aim to investigate the broader applicability of our \gram-based diversity framework beyond test generation, including its potential use in test selection, test prioritization, and for automated boundary value testing~\cite{dobslaw2023automated}.

\section{Data Availability}

Our replication package is publicly available~\cite{replication-package}, making our results reproducible.

\begin{acks}
This work was partially supported by the project Toposcope (SNF grant n. 214989). 
\end{acks}

\bibliographystyle{ACM-Reference-Format}
\bibliography{main}

\end{document}